\documentclass[conference]{IEEEtran}

\renewcommand\IEEEkeywordsname{Index Terms}
\usepackage{graphicx}
\usepackage{subcaption}
\usepackage{tikz,xparse}

\usetikzlibrary{patterns}

\usetikzlibrary{dsp,chains}
\usetikzlibrary{matrix}
\usepackage{mathptmx}
\usepackage{verbatim}
\usepackage{calc}
\usepackage{ifthen}
\usepackage{xifthen}
\usepackage{cancel}
\usepackage{bm}
\usepackage{verbatim}
\usepackage{multirow}
\usepackage{cite}
\usepackage[hyphens]{url}
\usepackage[nolist]{acronym} 
\usepackage{pgfplots}
\usetikzlibrary{arrows,shapes,graphs,graphs.standard,quotes,arrows.meta,decorations.markings}
\usetikzlibrary{matrix}
\pgfplotsset{compat=newest}
\usepackage[bookmarks=false]{hyperref}
\usepackage{units}
\usepackage{amsmath, amsbsy, amssymb, latexsym }
\hypersetup{bookmarksdepth=-2}
\usepackage{comment}
\usepackage[utf8]{inputenc}
\usepackage{xcolor}
\usepackage{enumitem}
\usepackage{algorithm}
\usepackage{algpseudocode}
\usetikzlibrary{spy}

\usepackage{marginnote}

\tikzset{>=latex}

\algnewcommand\algorithmicforeach{\textbf{for each}}
\algdef{S}[FOR]{ForEach}[1]{\algorithmicforeach\ #1\ \algorithmicdo}

\algnewcommand\algorithmicswitch{\textbf{switch}}
\algnewcommand\algorithmiccase{\textbf{case}}
\algnewcommand\algorithmicassert{\texttt{assert}}
\algnewcommand\Assert[1]{\State \algorithmicassert(#1)}%
\algdef{SE}[SWITCH]{Switch}{EndSwitch}[1]{\algorithmicswitch\ #1\ \algorithmicdo}{\algorithmicend\ \algorithmicswitch}%
\algdef{SE}[CASE]{Case}{EndCase}[1]{\algorithmiccase\ #1}{\algorithmicend\ \algorithmiccase}%
\algtext*{EndCase}%

\definecolor{mittelblau}{RGB}{0, 126, 198}
\definecolor{violettblau}{cmyk}{0.9, 0.6, 0, 0}
\definecolor{rot}{RGB}{238, 28 35}
\definecolor{apfelgruen}{RGB}{140, 198, 62}
\definecolor{gelb}{RGB}{255, 229, 0}
\definecolor{orange}{RGB}{244, 111, 33}
\definecolor{pink}{RGB}{237, 0, 140}
\definecolor{lila}{RGB}{128, 10, 145}
\definecolor{hellgrau}{RGB}{224, 224, 224}
\definecolor{mittelgrau}{RGB}{128, 128, 128}
\definecolor{dunkelgrau}{RGB}{80,80,80}
\definecolor{anthrazit}{RGB}{19, 31, 31}
\definecolor{darkgreen}{RGB}{34,139,34}
\definecolor{aqua}{RGB}{0, 255, 255}
\colorlet{Mycolor1}{green!10!orange!90!}
\tikzset{
       vnd/.style={
        shape=circle,
        fill=black,
        draw,
        inner sep=0pt,
        minimum size=0.2cm},
        cnd/.style={
        shape=rectangle,
        fill=white,
        draw,
        minimum width=0.05mm,
        minimum height = 0.05mm}, 
         vndR/.style={
        shape=circle,
        fill=red,
        draw,
        inner sep=0pt,
        minimum size=0.2cm},
        cndR/.style={
        shape=rectangle,
        fill=white,
        draw=red,
        minimum width=0.05mm,
        minimum height = 0.05mm}
}

\DeclareMathOperator*\bigboxplus{\ensuremath{\boxplus}}
\newcommand{\qed}{\hfill\blacksquare}

\IEEEoverridecommandlockouts

\def \reviewmode{\if01}	

\renewcommand{\vec}[1]{\mathbf{#1}}

\newcommand{\bv}{\vec{b}}

\newcommand{\gv}{\vec{g}}

\newcommand{\uv}{\vec{u}}
\newcommand{\vv}{\vec{v}}

\newcommand{\xv}{\vec{x}}
\newcommand{\yv}{\vec{y}}
\newcommand{\zv}{\vec{z}}
\newcommand{\zerov}{\vec{0}}

\newcommand{\Am}{\vec{A}}

\newcommand{\Gm}{\vec{G}}
\newcommand{\Hm}{\vec{H}}

\newcommand{\Lm}{\vec{L}}

\newcommand{\Pm}{\vec{P}}

\newcommand{\Um}{\vec{U}}

\newcommand{\FF}{\mathbb{F}}

\newcommand{\LB}{\left(}
\newcommand{\RB}{\right)}
\newcommand{\LP}{\left\{}
\newcommand{\RP}{\right\}}

\newcommand{\sgn}{\mathop{\mathrm{sgn}}}

\newcommand{\argmin}{\mathop{\mathrm{argmin}}}
\newcommand{\argmax}{\mathop{\mathrm{argmax}}}

\begin{document}
	
\begin{NoHyper}
\title{Automorphism Ensemble Decoding of Reed--Muller Codes}

\iftrue \reviewmode
	\author{\IEEEauthorblockN{Author 1, Author 2, Author 3, Author 4 and Author 5}
	\IEEEauthorblockA{}}
\else
	\author{\IEEEauthorblockN{Marvin Geiselhart, Ahmed Elkelesh, Moustafa Ebada, Sebastian Cammerer and Stephan ten Brink}\thanks{
			 Marvin Geiselhart, Ahmed Elkelesh, Moustafa Ebada and Stephan ten Brink are with the Institute of Telecommunications, Pfaffenwaldring 47, University of Stuttgart, 70569 Stuttgart, Germany (e-mail: {geiselhart,elkelesh,ebada,tenbrink}@inue.uni-stuttgart.de).
			 
			Sebastian Cammerer is with NVIDIA, Fasanenstraße 81, 10623 Berlin (e-mail: scammerer@nvidia.com); the work was carried out while he was with the University of Stuttgart.
			
			Parts of this work have been accepted in the International Symposium on Topics in Coding 2021 (ISTC), Sep. 2021.}

	\IEEEauthorblockA{
}
}

\fi

\maketitle

\begin{acronym}
\acro{ECC}{error-correcting code}
\acro{HDD}{hard decision decoding}
\acro{SDD}{soft decision decoding}
\acro{ML}{maximum likelihood}
\acro{GPU}{graphical processing unit}
\acro{BP}{belief propagation}
\acro{BPL}{belief propagation list}
\acro{LDPC}{low-density parity-check}
\acro{HDPC}{high density parity check}
\acro{BER}{bit error rate}
\acro{SNR}{signal-to-noise-ratio}
\acro{BPSK}{binary phase shift keying}
\acro{AWGN}{additive white Gaussian noise}
\acro{MSE}{mean squared error}
\acro{LLR}{Log-likelihood ratio}
\acro{LUT}{look-up table}
\acro{MAP}{maximum a posteriori}
\acro{NE}{normalized error}
\acro{BLER}{block error rate}
\acro{PE}{processing elements}
\acro{SCL}{successive cancellation list}
\acro{SC}{successive cancellation}
\acro{SCAN}{soft cancellation}
\acro{BI-DMC}{Binary Input Discrete Memoryless Channel}
\acro{CRC}{cyclic redundancy check}
\acro{CA-SCL}{CRC-aided successive cancellation list}
\acro{BEC}{Binary Erasure Channel}
\acro{BSC}{Binary Symmetric Channel}
\acro{BCH}{Bose-Chaudhuri-Hocquenghem}
\acro{RM}{Reed--Muller}
\acro{RS}{Reed-Solomon}
\acro{SISO}{soft-in/soft-out}
\acro{PSCL}{partitioned successive cancellation list}
\acro{3GPP}{3rd Generation Partnership Project }
\acro{eMBB}{enhanced Mobile Broadband}
\acro{PCC}{parity-check concatenated}
\acro{CA-polar codes}{CRC-aided polar codes}
\acro{CN}{check node}
\acro{VN}{variable node}
\acro{PC}{parity-check}
\acro{GenAlg}{Genetic Algorithm}
\acro{AI}{Artificial Intelligence}
\acro{MC}{Monte Carlo}
\acro{CSI}{Channel State Information}
\acro{PSCL}{partitioned successive cancellation list}
\acro{OSD}{ordered statistic decoding}
\acro{MWPC-BP}{minimum-weight parity-check BP}
\acro{FFG}{Forney-style factor graph}
\acro{MBBP}{multiple-bases belief propagation}
\acro{NBP}{neural belief propagation}
\acro{URLLC}{ultra-reliable low-latency communications}
\acro{DMC}{discrete memoryless channel}
\acro{MSB}{most significant bit}
\acro{LSB}{least significant bit}
\acro{RPA}{recursive projection-aggregation}
\acro{SGD}{stochastic gradient descent}
\end{acronym}

\begin{abstract}
Reed--Muller (RM) codes are known for their good \ac{ML} performance in the short block-length regime. Despite being one of the oldest classes of channel codes, finding a low complexity soft-input decoding scheme is still an open problem. In this work, we present a versatile decoding architecture for RM codes based on their rich automorphism group. The decoding algorithm can be seen as a generalization of multiple-bases belief propagation (MBBP) and may use any polar or RM decoder as constituent decoders. We provide extensive error-rate performance simulations for successive cancellation (SC)-, SC-list (SCL)- and belief propagation (BP)-based constituent decoders. We furthermore compare our results to existing decoding schemes and report a near-ML performance for the RM(3,7)-code (e.g., $0.04$ dB away from the ML bound at BLER of $10^{-3}$) at a competitive computational cost. Moreover, we provide some insights into the automorphism subgroups of RM codes and SC decoding and, thereby, prove the theoretical limitations of this method with respect to polar codes.

\end{abstract}
\acresetall

\begin{IEEEkeywords}
Reed--Muller Codes, Polar Codes, Code Automorphisms, Successive Cancellation Decoding, Belief Propagation Decoding, List Decoding, Ensemble Decoding.
\end{IEEEkeywords}

\acresetall
\vspace{-0.2cm}
\section{Introduction}

The current trend of \ac{URLLC} applications has urged the need for efficient short length coding schemes in combination with the availability of efficient decoders.
Besides many other coding schemes, this has lead to the revival of one of the oldest error-correcting codes, namely \ac{RM} codes \cite{Muller1954,Reed1954} -- potentially also due to some existent similarities between \ac{RM} code and the newly developed family of polar codes \cite{ArikanMain, StolteRekursivPlotkin}. 
On the one hand, \ac{RM} codes, as an example of algebraic codes, are known to be capacity-achieving over the \ac{BEC} for a given rate \cite{Abbe_RM_BEC,RM_Capacity_BEC}. Moreover, and practically even more relevant, they enjoy an impressive error-rate performance under \ac{ML} decoding, which also holds in the short length regime.
To this extent, several decoding algorithms have been developed in the course of \ac{RM} decoding. On the other hand, to the best of our knowledge, there is still a lack of practical decoders that are characterized by near-ML performance \emph{and} feasible decoding complexity/latency.

\begin{figure} [t]
	\centering
	\resizebox{0.975\columnwidth}{!}{\begin{tikzpicture}
\tikzset{
edge/.style = {thick,black},
mydiamond/.style={draw, diamond, aspect=2.7,text width=1.75cm, inner sep=0pt,  fill=white!90!red},
BPrectangle/.style={rectangle, draw, minimum size=1.5cm, fill=white!90!gray},
intrect/.style={rectangle, draw, minimum size=1cm, fill=white!90!gray}
}

\tikzstyle{conn} = [-{Latex[length=2mm,width=2mm]}];

\node[draw,shape=circle, fill=black, inner sep=0pt,minimum size=0.15cm, label=below:{$\mathbf{y}$}] (Lch) at (2, 0.5) {};

\draw [edge, dotted] (-2,-0.5)--(-2,-1);
\draw [edge] (Lch)--(1.5,0.5)--(1.5,-0.5);
\draw [edge, dotted] (1.5,-0.5)--(1.5,-1);

\draw [edge, black,-, dotted] (-4.5,-0.5)--(-4.5,-1);

	\draw [edge, black,-, dotted] (.5,-0.5)--(.5,-1);
	
	\node[BPrectangle] (BP1) at (-2, 3) {SC / BP / SCL};
	\node[intrect] (int1) at (.5, 3) {$ \pi_1 $};
	\node[intrect] (deint1) at (-4.5, 3) {$ \pi_1^{-1} $};
	\draw [edge,conn] (Lch)--(1.5,0.5)--(1.5,3)--(int1);
	
	\node[BPrectangle] (BP2) at (-2, 1) {SC / BP / SCL};
	\node[intrect] (int2) at (.5, 1) {$ \pi_2 $};
	\node[intrect] (deint2) at (-4.5, 1) {$ \pi_2^{-1} $};
	\draw [edge,conn] (Lch)--(1.5,0.5)--(1.5,1)--(int2);
	
	\node[BPrectangle] (BP3) at (-2, -2) {SC / BP / SCL};
	\node[intrect] (int3) at (.5, -2) {$ \pi_M $};
	\node[intrect] (deint3) at (-4.5, -2) {$ \pi_M^{-1} $};
	\draw [edge,conn] (1.5,-1)--(1.5,-2)--(int3);
	
	\draw [edge,conn](int1) to node[above] {$\mathbf{y}'_1$} (BP1);
	\draw [edge,conn](int2) to node[above] {$\mathbf{y}'_2$} (BP2);
	\draw [edge,conn](int3) to node[above] {$\mathbf{y}'_M$} (BP3);
	
	\draw [edge,conn](BP1) to node[above] {$\hat{\mathbf{x}}'_1$} (deint1);
	\draw [edge,conn](BP2) to node[above] {$\hat{\mathbf{x}}'_2$} (deint2);
	\draw [edge,conn](BP3) to node[above] {$\hat{\mathbf{x}}'_M$} (deint3);
	
	\draw [edge,conn](deint1) to node[above] {$\hat{\mathbf{x}}_1$} (-6.5,3);
	\draw [edge,conn](deint2) to node[above] {$\hat{\mathbf{x}}_2$} (-6.5,1);
	\draw [edge,conn](deint3) to node[above] {$\hat{\mathbf{x}}_M$} (-6.5,-2);

	\node[] (int) at (-2,-3.8) {Bit (de-) interleavers};
	\draw [black,-, ] (int)--(.5,-2.6);
	\draw [black,-] (int)--(-4.5,-2.6);

\node[rectangle, draw, minimum width=6cm, minimum height=0.3cm, text height=0.3cm, text depth=0.3cm,text centered,rotate=90, fill=white!90!cyan] (decide) at (-7, 0.5) {$\hat{\mathbf{x}}=\underset{\hat{\mathbf{x}}_{j},j\in\left\{ 1,\dots,M\right\} }{\mathrm{argmin}}\left\Vert \mathbf{y}-\hat{\mathbf{x}}_{j}\right\Vert $};

\draw [edge,conn] (Lch)--(1.75,0.5)--(1.75,4)--(-7,4)to node[right] {$\mathbf{y}$}(decide);

\draw [edge,conn](decide) to node[above] {$\hat{\mathbf{x}}$} (-8.5,0.5);

\end{tikzpicture}}
	\caption{\footnotesize Block diagram of automorphism ensemble decoding. $ M $ constituent decoders are used.}	
	\label{fig:Block_Diag}
\end{figure}
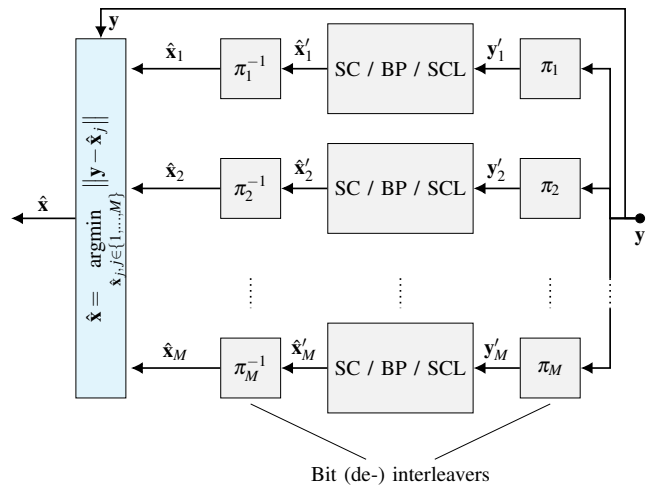

\ac{RM} decoders can be grouped into two main categories, iterative and non-iterative decoders. We revisit some of the famous non-iterative RM decoders.
In the literature, the best known decoder for RM codes over an \ac{AWGN} channel is Dumer’s recursive list decoding algorithm \cite{permuteRM}, which is now known under the name \ac{SCL} decoding, and a variant using permutations.
Recently, a \ac{RPA} decoding algorithm for RM codes was proposed in \cite{RPA_Abbe}, which can be viewed as a weighted BP decoder over a redundant factor graph \cite{WBP_RPA_Pfister_ISIT20}, making use of the symmetry of the RM codes (i.e., its large automorphism group).
RM codes under RPA decoding were shown to outperform the error-rate performance of CRC-aided polar codes under \ac{SCL} decoding.
It is worth mentioning that a more general usage of the rich automorphism group of RM codes (to aid the decoding process) is reported in \cite{StolteRekursivPlotkin}, along with the decoding of RM codes using a redundant parity-check matrix proposed earlier in \cite{Bossert_RM}.
In this work, the considered iterative decoders are the multiple-bases belief propagation\acused{MBBP} (\ac{MBBP}) decoding \cite{Huber}, minimum weight parity-check belief propagation\acused{MWPC-BP} (\ac{MWPC-BP}) decoding \cite{PfisterMWPC}, neural belief propagation\acused{NBP} (\ac{NBP}) decoding \cite{NachmaniNeuralBP}, pruned neural belief propagation (pruned-\ac{NBP}) decoding \cite{buchberger2020pruning} which is based on combining the idea of \ac{MWPC-BP} together with \ac{NBP}. These decoders will be introduced and briefly described in this paper while we focus on iterative decoding of \ac{RM} codes. The inherent parallel nature of iterative decoders allows high throughput implementations. Also, iterative decoders are \ac{SISO} decoders by its nature and, thus, are suitable for iterative detection and decoding \cite{ebada2020iterative}. 

Knowing that RM codes can be viewed, from an implementation perspective, as a polar code with a specific frozen/non-frozen bit pattern, we can decode RM codes with the recently proposed decoders for polar codes. Such decoders view the codes as codes on graphs \cite{Forney} by using the \ac{FFG} rather than a Tanner graph. Well known examples for these decoders are \ac{SC} \cite{ArikanMain}, \ac{BP} \cite{ArikanBP_original}, \ac{SCL} \cite{talvardyList} and \ac{SCAN} \cite{SCAN_Dec}. Furthermore, variants of these decoders were proposed to further enhance the error-correcting performance by using stage-permutations of the \ac{FFG} (e.g., \ac{BP} on permuted factor graphs \cite{multi_trellis}, \ac{BPL} \cite{elkelesh2018belief} and permuted SCL \cite{PermutedRM_Ivanov}).

In this paper, we propose a new decoding scheme, extending and generalizing some of the previously mentioned decoding algorithms. 
Fig.~\ref{fig:Block_Diag} shows an abstract view of our proposed decoding algorithm.
The main contributions of this paper are:

\begin{itemize}
	\item We present a new flexible framework for decoding \ac{RM} codes over the \ac{AWGN} channel. It is based on an ensemble of already existent polar decoders (e.g., \ac{SC}, \ac{BP} and \ac{SCL}) that run completely independently and in parallel. Each decoder instance uses a permutation from the rich automorphism group of the \ac{RM} code. Our framework can be seen as a generalization of the \ac{MBBP} decoder \cite{Huber}, the \ac{BPL} decoder \cite{elkelesh2018belief} and the decoder proposed in \cite{Moscow_Huawei_RM_paper}.
	\item We show that the automorphism group of \ac{RM} codes can be divided into smaller subgroups with different properties and investigate their performance in the proposed framework.\footnote{The source code of a sample implementation is provided online: \url{https://github.com/MGeiselhart/RM_AED}} By sampling from the full automorphism group, we outperform existing schemes that were restricted to specific subgroups (e.g., when compared to \cite{Huber,elkelesh2018belief,Moscow_Huawei_RM_paper,PfisterMWPC,buchberger2020pruning,NachmaniNeuralBP,talvardyList}).
	\item We provide extensive \ac{BLER} simulations and an operation-level complexity comparison of our scheme with existing decoding algorithms for RM codes.
	To the best of our knowledge, our proposed decoding algorithm achieves the best practical decoding performance of the RM(3,7)-code presented thus far (e.g., $0.04$ dB away from the ML bound at BLER of $10^{-3}$).
	\item Finally, we derive a property of the \ac{SC} decoder related to automorphism subgroups. We find that the permutations from the automorphism subgroup of general polar codes commute with the \ac{SC} decoding operation. This helps to explain why limitations of extending the algorithm to polar codes with \ac{SC} constituent decoders arise. These theoretical results are also backed-up by Monte-Carlo simulation.
\end{itemize}

The paper is organized as follows. 
In Sec.~\ref{sec:preliminaries}, we briefly review the concepts of RM codes, polar codes, code automorphisms and RM/polar decoding techniques.
In Sec~\ref{sec:ensemble_dec}, we introduce our proposed automorphism-based ensemble decoding algorithm.
We present some interesting facts about the automorphisms of the SC decoder in Sec.~\ref{sec:auto-SC}.
In Sec.~\ref{sec:Results}, we show error-rate performance and complexity comparison results for our proposed RM decoding scheme when compared to the state-of-the-art.
Sec.~\ref{sec:conc} renders some conclusions and opens up future work avenues.

\section{Preliminaries}\label{sec:preliminaries}
\subsection{Reed--Muller Codes}
\ac{RM} codes were first introduced in 1954 by David E. Muller as an algebraic coding scheme \cite{Muller1954} with their first efficient decoding algorithm  introduced in the same year by Irving Reed \cite{Reed1954}. The basic idea is to treat codewords as the evaluation of polynomials. For an \ac{RM} code, each message is interpreted as a multilinear polynomial $ u(\zv) $ in $ m $ binary variables $ z_i $, where $i\in\{0,\cdots,m-1\}$, and maximum degree $ r $ (usually called the order of the \ac{RM} code), over the finite field $ \FF_2 $. This can be written as

\begin{align}
u(\zv) &= u(z_0, \cdots , z_{m-1}) = \sum_{i=0}^{k-1} u_i \cdot g_i(\zv) = \uv\cdot \gv^T(\zv),
\end{align}
where $ \gv^T(\zv) $ denotes the vector of length $ k $ containing all monomials of maximum degree $ r $ in descending order. The number of such monomials is given as

\begin{equation}\label{eq:rm_k}
k = \sum_{i=0}^{r} \binom{m}{i} = \binom{m}{0} + \binom{m}{1} + \cdots + \binom{m}{r}.
\end{equation}

To obtain a codeword, the message polynomial is evaluated at all points in the space $ \FF_2^m $, resulting in $ N = 2^m $ codeword bits. Alternatively, the $ \gv^T(\zv) $ may be evaluated to form the generator matrix $ \Gm $:

\begin{align}\label{eq:poly_eval}
\xv &= \operatorname{eval}\left(u(\zv)\right) = \uv\cdot \underbrace{\operatorname{eval}\left(\gv^T(\zv)\right)}_{\Gm}= \uv \cdot \Gm.
\end{align}

To easily relate \ac{RM} and polar codes, we introduce this evaluation in reverse binary order\footnote{This does not change the code, as reversing the bit order is contained in the automorphism group of RM codes.}, i.e., the vector $ \zv $ is associated with the codeword bit $ x_i $ with $ i = \sum_{j=0}^{m-1}(1-z_j) \cdot 2^j $. That way, the generator matrix $ \Gm $ can be obtained by selecting those rows of the $N\times N$ Hadamard matrix $ \Gm_N = \left[ \begin{array}{ll} 1 & 0 \\ 1 & 1 \end{array}\right]^{\otimes m} $ that have an index with a Hamming weight of at least $ m-r $, where $(\cdot)^{\otimes m}$ denotes the $m$-th Kronecker power of a matrix \cite{Forney}. In the polar coding context, the position of these rows are referred to as the \emph{information bit positions}, while the position of the removed rows are the \emph{frozen bit positions}.

\subsection{Polar Codes}
The cornerstone upon which polar codes are constructed is the channel polarization concept, where, initially, $N$ copies of a \ac{DMC} are converted by recursive application of a \emph{channel transform} into $N$ synthetic channels. Those synthetic channels show a polarization behavior in being either sufficiently good (i.e., noiseless) to hold the information bits for transmission, or sufficiently poor (i.e., noisy) to not carrying any information at all and, thus, set to a fixed value known for both transmitter and receiver, hence \emph{frozen}. The above statement only holds strictly as the code length $N=2^m$ gets larger assuming \ac{SC} decoding.
Throughout this work, we set the value of the frozen-bit positions to ``0''.

Without loss of generality, a \emph{polar code} refers to the set of synthetic bit-channels used for information transmission or, equivalently, their complementary set of bit-channels set to be frozen, denoted by the information set $\mathbb{A}$ and the frozen set $\bar{\mathbb{A}}$, respectively. Eventually, the indices of the information set refers to the rows selected from $\mathbf{G}_N$ (which is also called the \emph{polarization matrix} in the context of polar codes), that constitute the $k\times N$ polar code generator matrix.

According to this view of polar codes, \ac{RM} and polar codes differ only in the row-selection criterion from $\mathbf{G}_N$. Therefore one can infer that the row-selection criterion of the \ac{RM} codes optimizes the code performance under \ac{MAP} decoding, whereas that of polar codes optimizes the code performance under \ac{SC} decoding. For that, \ac{RM} codes are characterized by a better \ac{MAP} threshold when compared to polar codes, while, however, decoders that manage to approach the RM code's MAP threshold are of impractical complexity, which leaves the door open for further research in finding a practical \ac{RM} decoding strategy that achieves the RM code's MAP performance \cite{RMurbankePolar}.

In the same way as \ac{RM} codes, polar codes can be viewed in terms of evaluation of polynomials \cite{bardet_polar_automorphism}. There is a one-to-one mapping from rows $ i $ of $ \Gm_N $ to monomials in $ m $ variables given by

\begin{equation}
 i \mapsto \prod_{j=0}^{m-1} (z_j)^{1 - i_j},
\end{equation}
where $ i_j $ denotes the $ j $-th bit in the binary representation of $ i $, i.e., $ i=\sum_{j=0}^{m-1} i_j \cdot 2^j $. A polar code is then defined according to Eq. (\ref{eq:poly_eval}) by a monomial vector $ \gv^T(\zv) $ that contains the monomials corresponding to the information set $ \mathbb{A} $. We further define $ \mathcal{M}_m $ as the set of all monomials in $ m $ variables and denote by $ I~=~\LP {g_i(\zv) }\RP \subseteq \mathcal{M}_m $ the (unordered) set of monomials of a polar code. Hence, $ \mathbb{A} $, $ I $ and $ \gv(\zv) $ are all equivalent representations of the same polar code.

Moreover, most practical polar code designs fulfill the so-called partial ordering of the monomials (and therefore synthetic channels) making them \emph{decreasing monomial codes}. These codes are the only group of polar codes where algebraic properties are known and therefore of particular value for algebraic decoding.
We quickly recapitulate the definition of decreasing monomial codes as given in \cite{bardet_polar_automorphism}:

\textbf{Definition (Decreasing Monomial Code):} A polar code with monomial set $ I $ is said to be a \emph{decreasing monomial code}, if

\begin{equation}\label{def:dec_mon_code}
\forall g \in I, \forall f \in \mathcal{M}_m \text{ with } f \preccurlyeq g \Rightarrow f\in I
\end{equation}
holds, i.e., if a synthetic channel carries information, then all \emph{better} channels (according to the partial order) must also be nonfrozen. This partial order `$ \preccurlyeq $' of monomials is defined as

\begin{equation}\label{def:partial_order1}
 z_{i_1}\cdots z_{i_s}\preccurlyeq z_{j_1}\cdots z_{j_s} \Leftrightarrow i_k \le j_k \forall k 
\end{equation}
for monomials of equal degree and
\begin{align}\label{def:partial_order2}
f \preccurlyeq g \Leftrightarrow \exists g^* | g &\text{ with } \operatorname{deg}(g^*) = \operatorname{deg}(f) \text{ and } f \preccurlyeq g^*
\end{align}
if $ \operatorname{deg}(f) < \operatorname{deg}(g) $. For more details on decreasing monomial codes, we refer the interested reader to \cite{bardet_polar_automorphism}.

\subsection{Automorphism Group}
The \emph{automorphism group} (or \emph{permutation group}) $ \operatorname{Aut}(\mathcal{C}) $ of a code $ \mathcal{C} $ is the set of permutations $ \pi $ of the codeword bit indices that map $ \mathcal{C} $ onto itself, i.e. 

\begin{equation}
	\pi(\xv) \in \mathcal{C} \quad \forall \xv \in \mathcal{C} \quad \forall \pi \in \operatorname{Aut}(\mathcal{C}),
\end{equation}
where $ \pi(\xv) $ results in the vector $ \xv' $ with $ x_i' = x_{\pi(i)} $. In other words, every codeword is mapped to another (\emph{not} necessarily different) codeword of the same code. $\operatorname{Aut}(\mathcal{C}) $ forms a subgroup of the symmetric group $ \mathcal{S}_N $ under permutation composition \cite{macwilliams77}.

\subsubsection{RM Code Automorphisms}
The automorphism group of \ac{RM} codes is well known as the \emph{general affine group} GA($ m $) over the field $ \FF_2 $\cite{macwilliams77}.\footnote{In this paper, we only consider the field $ \FF_2 $ and hence, we omit the size of the field in the notation, i.e., we write GA($ m $) instead of GA($ m,2 $).} GA($ m $) is the group of all affine bijections over $ \FF_2^m $, i.e., pairs $ (\Am, \bv) $ defining the mapping $ \zv' = \Am \zv + \bv $, with a non-singular matrix $ \Am \in \FF_2^{m\times m}$ and an arbitrary vector $ \bv \in \FF_2^{m\times 1} $. The vectors $ \zv, \zv' \in \FF_2^{m\times 1} $ are the binary representations of the code bit positions $ i $ and $ \pi(i) $, respectively, i.e., $ i = \sum_{j=0}^{m-1}{z_j \cdot 2^j} $. In the following, we will use the permutation notation $ i' = \pi(i) $ and its matrix-vector pair $ (\Am, \bv) $ interchangeably.

\subsubsection{Polar Code Automorphisms}
To this point, the automorphism group of general polar codes is unknown. However, for decreasing monomial codes (i.e., practically relevant polar codes), a subset of the automorphism group is known to be the lower-triangular affine group LTA($ m $). It is characterized by pairs $ (\Am, \bv) $, where  $ \Am \in \FF_2^{m\times m}$ is a lower-triangular matrix with a unit diagonal and arbitrary $ \bv \in \FF_2^{m\times 1} $ \cite{bardet_polar_automorphism}. The underlying permutation of codeword bit indices works identical to \ac{RM} codes. It is easy to see that LTA($ m $) is a proper subgroup of GA($ m $). Therefore, as polar codes are a generalization of \ac{RM} codes, they have generally fewer automorphisms. It should be emphasized that LTA($ m $) constitutes only a part of the automorphisms of polar codes and, depending on the polar code design (i.e., the information set $ \mathbb{A} $), more permutations may be part of the full automorphism group.

\subsubsection{Automorphism Subgroups}\label{sec:subgroups}
Similar to $ \operatorname{LTA}(m) $, we can define the \emph{upper-triangular affine group} $ \operatorname{UTA}(m) $ as the subgroup of $ \operatorname{GA}(m) $ with $ \Am $ being upper-triangular with a unit diagonal. Moreover, we may define the subgroup $ \Pi(m) $, where $\Am$ is a permutation matrix and $ \bv = \zerov $, corresponding to the \emph{stage-shuffle permutations} of the \ac{FFG}.
Obviously, $ \operatorname{LTA}(m) $, $ \operatorname{UTA}(m) $ and $ \Pi(m) $ are all subgroups of $ \operatorname{GA}(m) $, because their $ \Am $-matrices and $ \bv $-vectors are special cases of the arbitrary non-singular $ \Am \in \FF_2^{m\times m} $ and $ \bv \in \FF_2^{m\times 1} $. Moreover, the sets are closed under composition `$\circ$' and their inverses exist, as triangular and permutation matrices are always invertible.

\textbf{Theorem 1}:
Every permutation $ \pi \in \operatorname{GA}(m) $ can be written as a composition $ \pi_L \circ \pi_U \circ \pi_P $, with  $ \pi_L~\in~\operatorname{LTA}(m) $, $ \pi_U~\in~\operatorname{UTA}(m) $ and $ \pi_P~\in~\Pi(m) $.

\emph{Proof:}
We first establish that any non-singular square matrix $ \Am $ can be factorized as $ \Am = \tilde{\Lm}\tilde{\Um}\tilde{\Pm} $, where $ \tilde{\Lm} $ is lower triangular, $ \tilde{\Um} $ upper triangular and $ \tilde{\Pm} $ is a permutation matrix; we call this the \emph{modified LUP decomposition}. It can be derived from the well-known LUP decomposition \cite[Chapter~31]{cormen_algorithms_2001} of the matrix $\Am^T$ as
\begin{align}
\Pm \Am^T &= \Lm\Um \\
\Am^T &= \Pm^{-1} \Lm \Um \\
\Am &= (\Pm^{-1} \Lm \Um)^T = \Um^T\Lm^T(\Pm^{-1})^T = \tilde{\Lm}\tilde{\Um}\tilde{\Pm},
\end{align}
with $ \tilde{\Lm} = \Um^T $, $ \tilde{\Um} = \Lm^T $ and $ \tilde{\Pm} = (\Pm^{-1})^T $. We can use this fact to prove the original statement as
\begin{align}
i' = \pi(i) \Leftrightarrow \zv' &= \Am \zv + \bv \nonumber\\
&=(\Am_L \Am_U \Am_P) \zv + \bv \nonumber\\
&=\Am_L \left( \Am_U \Am_P \zv + \bv_U \right) + \bv_L \\
\Leftrightarrow i' = \pi_L \left( \pi_U \left(\pi_P(i)\right)\right) &= (\pi_L \circ \pi_U \circ \pi_P)(i),
\end{align}
with $\Am_L$, $\Am_U$ and $\Am_P$ are obtained from the modified LUP decomposition of $ \Am $ and $ \bv_U=\zerov $ and $ \bv_L=\bv $. $\qed$

\textbf{Corollary 1.1}: The general affine group $ \operatorname{GA}(m) $ is generated from the elements of the union of the lower-triangular affine group $ \operatorname{LTA}(m) $, the upper-triangular affine group $\operatorname{UTA}(m) $ and stage-shuffle permutations $ \Pi(m) $.\footnote{Strictly speaking, $ \operatorname{UTA}(m) $ is not necessary in this statement, as it is generated from $\operatorname{LTA}(m) $ and $ \Pi(m) $, by observing that a lower triangular matrix can be transformed into an upper triangular matrix by reversing the order of the rows and columns. For simplicity, we only consider the decomposition into at most three different permutations, which requires the use of $ \operatorname{UTA}(m) $.}

\emph{Proof:}
From the subgroup property, the union of $ \operatorname{LTA}(m) $, $ \operatorname{UTA}(m) $ and $ \Pi(m) $ cannot generate elements outside $ \operatorname{GA}(m) $. From Theorem 1 we know that at least all elements in $ \operatorname{GA}(m) $ may be generated. Hence, $ \operatorname{LTA}(m) $, $ \operatorname{UTA}(m) $ and $ \Pi(m) $ exactly generate  $ \operatorname{GA}(m) $. $\qed$

\subsection{Polar and RM Decoding}
In this section, we briefly revise the different decoding techniques which can be used for \ac{RM} codes.

\subsubsection{SC Decoding}

\begin{figure}[t]
	\centering{\begin{tikzpicture}
	\tikzstyle{arrow} = [dspconn,line width = 0.25mm];
	\tikzstyle{normalline} = [line width = 0.5mm];
	\tikzstyle{block} = [rectangle, draw, minimum size=1.5cm, fill=white!90!gray],

	\node[] (p1) at (-3.7, -2) {}; 
	\node[] (p4) at (.7, 2) {};
	\filldraw [fill=gray!20!white,draw=gray,dashed,line width = 0.3mm] (p1) rectangle (p4);
	\node[scale=1] (text) at (-1.5, 2.2) {$ \operatorname{SC}_s $};
	
	\node[block] (SC1) at (-2.75, 1) {$ \operatorname{SC}_{s-1} $};
	\node[block] (SC2) at (-2.75, -1) {$ \operatorname{SC}_{s-1} $};
	
	\node[dspnodefull,minimum size=1.5mm] (temp) at (0, -1) {};
	\node[dspadder](xor) at (0, 1) {};
	\node[] (x1) at (2, 1) {};
	\node[] (x2) at (2, -1) {};
	\draw[normalline](SC1)--(xor)--(x1);
	\draw[normalline](SC2)--(temp)--(x2);
	\draw[normalline](xor)--(temp);

	\draw[arrow]  (-.8,1.2) --node[above] {$L_{i,s}$}(-1.8,1.2);
	\draw[arrow]  (-1.8,.8) --node[below] {$u_{i,s}$}(-.8,0.8);
	
	\draw[arrow]  (-.8,-0.8) --node[above] {$L_{i+2^s,s}$}(-1.8,-0.8);
	\draw[arrow]  (-1.8,-1.2)--node[below] {$u_{i+2^s,s}$}(-.8,-1.2);
	
	\draw[arrow]  (1.8,1.2) --node[above] {$L_{i,s+1}$}(.8,1.2);
	\draw[arrow]  (.8,.8)   --node[below] {$u_{i,s+1}$}(1.8,0.8);
	
	\draw[arrow]  (1.8,-0.8) --node[above, xshift=1ex] {$L_{i+2^s,s+1}$}(.8,-0.8);
	\draw[arrow]  (.8,-1.2)  --node[below, xshift=1ex] {$u_{i+2^s,s+1}$}(1.8,-1.2);
						
\end{tikzpicture}}
	\caption{\footnotesize Block diagram of successive cancellation (SC) decoding at stage $s$ by using recursion.}
	\label{fig:sc_tree}
\end{figure}
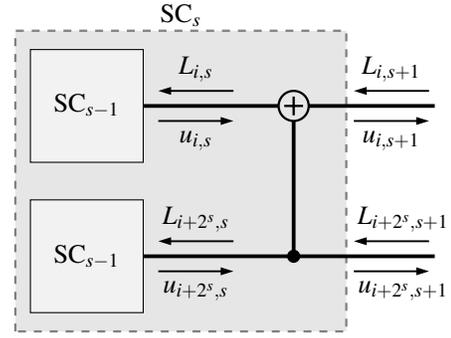

Polar codes and RM codes can be decoded by an \ac{SC} decoder where each information bit $u_i$ can be decoded via observing the channel noisy data $\mathbf{y}$ and the previously decoded bits $\hat{\mathbf{u}}_0^{i-1}$. Decoding complexity can be obviously preserved through imposing the code constraints via setting $u_i$ to be equal to the known frozen bit value of 0 for $i\in\bar{\mathbb{A}}$. 

Fig.~\ref{fig:sc_tree} shows the working principle of \ac{SC} decoding. Soft \ac{LLR} messages passes from right-to-left in a recursive manner on a binary tree representation of the \ac{FFG} of the \ac{RM} code. The right-most messages are given as the channel \acp{LLR}, i.e., $ L_{i,m} = L_{\mathrm{ch},i} $. The update rules for the messages to the first and second child nodes at each stage $ s $ are given by
\begin{align}
	L_{i,s}&=L_{i,s+1}\boxplus L_{i+2^{s},s+1},\label{eq:first_child_llr}\\
	L_{i+2^{s},s}&=(-1)^{u_{i,s}} \cdot L_{i,s+1} + L_{i+2^{s},s+1},\label{eq:second_child_llr}
\end{align} where $L_{i,s}$ is the soft \ac{LLR} value of the $i^{th}$ node at stage $s$, and the so-called \emph{box-plus} function is defined as

\begin{equation}\label{eq:boxplus}
a \boxplus b \triangleq\log\left(\dfrac{e^{a+b}+1}{e^{a}+e^{b}}\right).
\end{equation}
The recursion terminates at the left-most stage (i.e., in the leaf nodes) where either a hard decision on the \ac{LLR} is made (for information bits) or the frozen bit value 0 is returned (for frozen bits), which can be expressed as

\begin{align}
u_{i,0}&=\begin{cases}
0&i\in \bar{\mathbb{A}}\\
\operatorname{HD}(L_{i,0})&i\in \mathbb{A}\\
\end{cases}\label{eq:u_hard},\end{align}

with 

\begin{align}
\operatorname{HD}(L)\triangleq\begin{cases}
0&L\ge 0 \\
1&L < 0 \\
\end{cases}\label{eq:hard_dec}.\end{align}

The hard decisions are then propagated right according to

\begin{align}
u_{i,s+1}&=u_{i,s}\oplus u_{i+2^s,s},\label{eq:first_child_hard_dec}\\
u_{i+2^s,s+1}&=u_{i+2^s,s}.\label{eq:second_child_hard_dec}
\end{align}

Note that the equations impose a fixed order of computation, i.e., to compute Eq.~(\ref{eq:second_child_llr}) one has to first recurse Eq.~(\ref{eq:first_child_llr}) all the way to the very left, perform the hard decision Eq.~(\ref{eq:u_hard}) and propagate $ u_{i,s} $ according to Eq.~(\ref{eq:first_child_hard_dec}) back to the right. Finally, the message estimate and the codeword estimate of \ac{SC} decoding is given by the left and right-most hard-decisions, respectively, i.e., $ \hat{\uv} = \uv_0 $ and $ \hat{\xv} = \uv_m $. While being an asymptotically optimal decoding scheme for long polar codes, \ac{SC} decoding of \ac{RM} codes suffers from a poor error-rate performance which limits its practical use as a standalone decoder for \ac{RM} codes.

\subsubsection{SCL Decoding}
Instead of the one-branch \ac{SC} decoding where bits are hard-decided in a sequential manner affecting all decoding decisions yet-to-come, the breadth-first search \ac{SCL} decoding strategy branches out while decoding, continuing in both possible values of each bit $u_i$ in a soft manner. The exponential growth of this search process is restricted by the predefined list size $L$, which defines the maximum number of parallel branches considered. Path metrics are then assigned to each single decoding path and used to truncate the list of branches after reaching that limit, only keeping the $L$ most promising candidates. 
It was shown in \cite{talvardyList} that a list size of practical complexity was sufficient for polar codes to work close to its \ac{ML} performance under \ac{SCL} decoding. However, for \ac{RM} codes, impractically large list sizes are required to perform close-to-optimum under \ac{SCL} decoding \cite{PermutedSCL}. 
A variant to \ac{SCL} decoding is proposed in \cite{permuteRM}, where the decoding is started with a list of permuted received LLRs using a subset of stage shuffle permutations. We refer to this variant as Dumer-Shabunov (DS) decoding.

\subsubsection{Belief Propagation Decoding over Forney-style Factor Graph}\label{sec:ffg}
Rather than on a Tanner graph, \ac{BP} decoding can also be performed over a \ac{FFG}, constructed from check and variable nodes of degree three \cite{Forney,ArikanBP_original}. Fig.~\ref{fig:rm_ffg} shows the \ac{FFG} of the RM(1,3)-code. The channel output \acp{LLR} are fed to the right-most nodes of the \ac{FFG}. The frozen bits are known to be 0 and, thus, contribute a priori \acp{LLR} of $ +\infty $ to the left-most nodes in the factor graph. \ac{LLR} messages propagate from right-to-left and then from left-to-right until a maximum user defined number of iterations $N_{\mathrm{it,max}}$ is reached, or a certain early stopping condition is satisfied.\footnote{Throughout this work, we use a $\Gm$-matrix-based stopping condition (i.e., stop when $\hat{\xv} = \hat{\uv} \cdot \Gm$).} A hard decision is applied to estimate the information bits $\hat{\uv}$ (left-most nodes) and the codeword bits $\hat{\xv}$ (right-most nodes). For more implementation details we refer the interested reader to~\cite{ArikanBP_original}.

As the frozen nodes always contribute the same \acp{LLR} to the equations, the \ac{FFG} can be reduced as shown on the right in Fig.~\ref{fig:rm_ffg}, by removing edges of constant value. Dashed lines indicate edges which are only computed in the right-to-left pass in order to estimate $ \hat{\uv} $. This potentially reduces the number of performed arithmetic operations per iteration while preserving the same performance in terms of error-rate \cite{Forney}. Throughout this work, whenever the BP decoding is used over the \ac{FFG}, we use the reduced version.

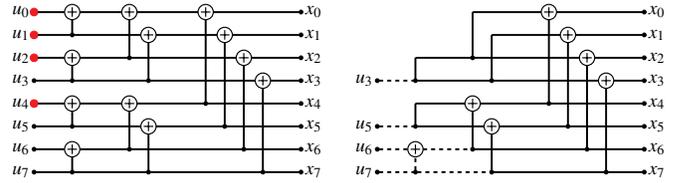
\begin{figure}[t]
	\centering\resizebox{\columnwidth}{!}{\begin{tikzpicture}[yscale=.6]
\tikzstyle{every node}=[font=\Large]
\tikzstyle{frozennode} = [dspnodefull,minimum size=2mm,rot];
\tikzstyle{normalnode} = [dspnodefull,minimum size=1mm];
\tikzstyle{normalline} = [line width = 0.5mm];
\tikzstyle{halfline} = [line width = 0.5mm, dashed];
\tikzstyle{conn} = [dspconn, line width = 0.5mm];

\node[frozennode] (u0) at (0.00, 7) {};
\node[left] at (0.00, 7) {$ u_{0} $};
\node[frozennode] (u1) at (0.00, 6) {};
\node[left] at (0.00, 6) {$ u_{1} $};
\node[frozennode] (u2) at (0.00, 5) {};
\node[left] at (0.00, 5) {$ u_{2} $};
\node[normalnode] (u3) at (0.00, 4) {};
\node[left] at (0.00, 4) {$ u_{3} $};
\node[frozennode] (u4) at (0.00, 3) {};
\node[left] at (0.00, 3) {$ u_{4} $};
\node[normalnode] (u5) at (0.00, 2) {};
\node[left] at (0.00, 2) {$ u_{5} $};
\node[normalnode] (u6) at (0.00, 1) {};
\node[left] at (0.00, 1) {$ u_{6} $};
\node[normalnode] (u7) at (0.00, 0) {};
\node[left] at (0.00, 0) {$ u_{7} $};
\node[dspadder] (node0) at (1.00, 7) {};
\node[normalnode] (node1) at (1.00, 6) {};
\draw[normalline] (u0)--(node0);
\draw[normalline] (u1)--(node1);
\draw[normalline] (node1)--(node0);
\node[dspadder] (node2) at (1.00, 5) {};
\node[normalnode] (node3) at (1.00, 4) {};
\draw[normalline] (u2)--(node2);
\draw[normalline] (u3)--(node3);
\draw[normalline] (node3)--(node2);
\node[dspadder] (node4) at (1.00, 3) {};
\node[normalnode] (node5) at (1.00, 2) {};
\draw[normalline] (u4)--(node4);
\draw[normalline] (u5)--(node5);
\draw[normalline] (node5)--(node4);
\node[dspadder] (node6) at (1.00, 1) {};
\node[normalnode] (node7) at (1.00, 0) {};
\draw[normalline] (u6)--(node6);
\draw[normalline] (u7)--(node7);
\draw[normalline] (node7)--(node6);
\node[dspadder] (node8) at (2.50, 7) {};
\node[normalnode] (node9) at (2.50, 5) {};
\draw[normalline] (node0)--(node8);
\draw[normalline] (node2)--(node9);
\draw[normalline] (node9)--(node8);
\node[dspadder] (node10) at (3.00, 6) {};
\node[normalnode] (node11) at (3.00, 4) {};
\draw[normalline] (node1)--(node10);
\draw[normalline] (node3)--(node11);
\draw[normalline] (node11)--(node10);
\node[dspadder] (node12) at (2.50, 3) {};
\node[normalnode] (node13) at (2.50, 1) {};
\draw[normalline] (node4)--(node12);
\draw[normalline] (node6)--(node13);
\draw[normalline] (node13)--(node12);
\node[dspadder] (node14) at (3.00, 2) {};
\node[normalnode] (node15) at (3.00, 0) {};
\draw[normalline] (node5)--(node14);
\draw[normalline] (node7)--(node15);
\draw[normalline] (node15)--(node14);
\node[dspadder] (node16) at (4.50, 7) {};
\node[normalnode] (node17) at (4.50, 3) {};
\draw[normalline] (node8)--(node16);
\draw[normalline] (node12)--(node17);
\draw[normalline] (node17)--(node16);
\node[dspadder] (node18) at (5.00, 6) {};
\node[normalnode] (node19) at (5.00, 2) {};
\draw[normalline] (node10)--(node18);
\draw[normalline] (node14)--(node19);
\draw[normalline] (node19)--(node18);
\node[dspadder] (node20) at (5.50, 5) {};
\node[normalnode] (node21) at (5.50, 1) {};
\draw[normalline] (node9)--(node20);
\draw[normalline] (node13)--(node21);
\draw[normalline] (node21)--(node20);
\node[dspadder] (node22) at (6.00, 4) {};
\node[normalnode] (node23) at (6.00, 0) {};
\draw[normalline] (node11)--(node22);
\draw[normalline] (node15)--(node23);
\draw[normalline] (node23)--(node22);
\node[normalnode] (x0) at (7.00, 7) {};
\node[right] at (7.00, 7) {$ x_{0} $};
\draw[normalline] (node16)--(x0);
\node[normalnode] (x1) at (7.00, 6) {};
\node[right] at (7.00, 6) {$ x_{1} $};
\draw[normalline] (node18)--(x1);
\node[normalnode] (x2) at (7.00, 5) {};
\node[right] at (7.00, 5) {$ x_{2} $};
\draw[normalline] (node20)--(x2);
\node[normalnode] (x3) at (7.00, 4) {};
\node[right] at (7.00, 4) {$ x_{3} $};
\draw[normalline] (node22)--(x3);
\node[normalnode] (x4) at (7.00, 3) {};
\node[right] at (7.00, 3) {$ x_{4} $};
\draw[normalline] (node17)--(x4);
\node[normalnode] (x5) at (7.00, 2) {};
\node[right] at (7.00, 2) {$ x_{5} $};
\draw[normalline] (node19)--(x5);
\node[normalnode] (x6) at (7.00, 1) {};
\node[right] at (7.00, 1) {$ x_{6} $};
\draw[normalline] (node21)--(x6);
\node[normalnode] (x7) at (7.00, 0) {};
\node[right] at (7.00, 0) {$ x_{7} $};
\draw[normalline] (node23)--(x7);

\node[normalnode] (u3) at (9.00, 4) {};
\node[left] at (9.00, 4) {$ u_{3} $};
\node[normalnode] (u5) at (9.00, 2) {};
\node[left] at (9.00, 2) {$ u_{5} $};
\node[normalnode] (u6) at (9.00, 1) {};
\node[left] at (9.00, 1) {$ u_{6} $};
\node[normalnode] (u7) at (9.00, 0) {};
\node[left] at (9.00, 0) {$ u_{7} $};
\coordinate (node2) at (10.00, 5) {};
\node[normalnode] (node3) at (10.00, 4) {};
\draw[halfline] (u3)--(node3);
\draw[normalline] (node3)--(node2);
\coordinate (node4) at (10.00, 3) {};
\node[normalnode] (node5) at (10.00, 2) {};
\draw[halfline] (u5)--(node5);
\draw[normalline] (node5)--(node4);
\node[dspadder] (node6) at (10.00, 1) {};
\node[normalnode] (node7) at (10.00, 0) {};
\draw[halfline] (u6)--(node6);
\draw[halfline] (u7)--(node7);
\draw[halfline] (node7)--(node6);
\coordinate (node8) at (11.50, 7) {};
\node[normalnode] (node9) at (11.50, 5) {};
\draw[normalline] (node2)--(node9);
\draw[normalline] (node9)--(node8);
\coordinate (node10) at (12.00, 6) {};
\node[normalnode] (node11) at (12.00, 4) {};
\draw[normalline] (node3)--(node11);
\draw[normalline] (node11)--(node10);
\node[dspadder] (node12) at (11.50, 3) {};
\node[normalnode] (node13) at (11.50, 1) {};
\draw[normalline] (node4)--(node12);
\draw[halfline] (node6)--(node13);
\draw[normalline] (node13)--(node12);
\node[dspadder] (node14) at (12.00, 2) {};
\node[normalnode] (node15) at (12.00, 0) {};
\draw[normalline] (node5)--(node14);
\draw[halfline] (node7)--(node15);
\draw[normalline] (node15)--(node14);
\node[dspadder] (node16) at (13.50, 7) {};
\node[normalnode] (node17) at (13.50, 3) {};
\draw[normalline] (node8)--(node16);
\draw[normalline] (node12)--(node17);
\draw[normalline] (node17)--(node16);
\node[dspadder] (node18) at (14.00, 6) {};
\node[normalnode] (node19) at (14.00, 2) {};
\draw[normalline] (node10)--(node18);
\draw[normalline] (node14)--(node19);
\draw[normalline] (node19)--(node18);
\node[dspadder] (node20) at (14.50, 5) {};
\node[normalnode] (node21) at (14.50, 1) {};
\draw[normalline] (node9)--(node20);
\draw[normalline] (node13)--(node21);
\draw[normalline] (node21)--(node20);
\node[dspadder] (node22) at (15.00, 4) {};
\node[normalnode] (node23) at (15.00, 0) {};
\draw[normalline] (node11)--(node22);
\draw[normalline] (node15)--(node23);
\draw[normalline] (node23)--(node22);
\node[normalnode] (x0) at (16.00, 7) {};
\node[right] at (16.00, 7) {$ x_{0} $};
\draw[normalline] (node16)--(x0);
\node[normalnode] (x1) at (16.00, 6) {};
\node[right] at (16.00, 6) {$ x_{1} $};
\draw[normalline] (node18)--(x1);
\node[normalnode] (x2) at (16.00, 5) {};
\node[right] at (16.00, 5) {$ x_{2} $};
\draw[normalline] (node20)--(x2);
\node[normalnode] (x3) at (16.00, 4) {};
\node[right] at (16.00, 4) {$ x_{3} $};
\draw[normalline] (node22)--(x3);
\node[normalnode] (x4) at (16.00, 3) {};
\node[right] at (16.00, 3) {$ x_{4} $};
\draw[normalline] (node17)--(x4);
\node[normalnode] (x5) at (16.00, 2) {};
\node[right] at (16.00, 2) {$ x_{5} $};
\draw[normalline] (node19)--(x5);
\node[normalnode] (x6) at (16.00, 1) {};
\node[right] at (16.00, 1) {$ x_{6} $};
\draw[normalline] (node21)--(x6);
\node[normalnode] (x7) at (16.00, 0) {};
\node[right] at (16.00, 0) {$ x_{7} $};
\draw[normalline] (node23)--(x7);

\end{tikzpicture}}
	\vspace{-0.5cm}
	\caption{\footnotesize Forney-style factor graph (FFG) of the RM(1,3)-code (left) and the reduced FFG (right). Note that the dashed lines indicate variables to be computed only in the right-to-left message update.}
	\label{fig:rm_ffg}
	\vspace{-0.5cm}	
\end{figure}

\subsubsection{Na\"ive Belief Propagation Decoding}
	\ac{BP} is a well-known decoding method for \ac{LDPC} codes. It is based on message passing on the Tanner graph of the parity-check matrix. Using its duality property, the parity-check matrix of \ac{RM} codes is found as the generator matrix of the dual \ac{RM} code. \ac{BP} decoding can be performed over the Tanner graph of this na\"ive parity-check matrix. However, the performance of this decoder is poor due to the numerous cycles in the graph induced by the high density of the parity-check matrix.
	
	\subsubsection{Minimum Weight Parity-Check Belief Propagation Decoding}
	Minimum weight parity-check belief propagation\acused{MWPC-BP} (\ac{MWPC-BP}) decoding introduced in \cite{PfisterMWPC} is based on the concept of iterative decoding over an overcomplete parity-check matrix. An online algorithm tailored to the noisy received sequence $ \yv $ is used to construct the overcomplete parity-check matrix only based on minimum weight parity-checks. These are found as the minimum weight codewords of the dual \ac{RM} code. Additionally, an attenuation factor for all check-to-variable node messages is applied to mitigate short cycle effects.
	
	\subsubsection{Neural Belief Propagation Decoding}
	Neural belief propagation\acused{NBP} (\ac{NBP}) decoding as introduced in \cite{NachmaniNeuralBP} treats the unrolled Tanner graph of the code as a neural network (NN), while assigning \emph{trainable} weights to all of its edges leading to a \emph{soft} Tanner graph. These trainable weights are optimized based on \ac{SGD} technique. The intuition behind why this algorithm enhances the error-rate performance even for dense factor graphs, is that the effect of graph cycles can be mitigated by the learned weights per edges over the whole graph. 
	
	\subsubsection{Pruned Neural Belief Propagation Decoding}
	Pruned neural belief propagation (pruned-\ac{NBP}) decoding \cite{buchberger2020pruning} combines the idea of \ac{MWPC-BP} together with \ac{NBP}. To get started, a redundant parity-check matrix containing (all or some of) the minimum weight parity-checks is constructed. During the offline training phase, all edges connected to a check node are assigned a single trainable weight and the least effective (i.e., least contributing) check node is \emph{pruned} (i.e., removed) from the graph. The authors of \cite{buchberger2020pruning} refer to this decoder as $ D_1 $. The error-rate performance of this algorithm can be further enhanced by assigning trainable weights per edge at the expense of larger memory requirements to save all weights per edges, resulting in decoder $ D_3 $. Furthermore, a pruned \ac{NBP} decoder without any weights is introduced as $ D_2 $, however with the expense of a significant degradation in error-rate performance.

\section{Automorphism Ensemble Decoding}\label{sec:ensemble_dec}
Ensemble decoding uses multiple constituent decoders (i.e., a decoder \emph{ensemble} of size $ M $) to generate a set of codeword estimates; and selects one of these codewords, using a predefined metric, as the decoder output. Typically, a least-squares metric is used, as this corresponds to the \ac{ML} decision for the \ac{AWGN} channel. Hence, this method is also called \emph{ML-in-the-list}, as it selects the optimal candidate from the list of codeword estimates. This can be formulated as
\begin{equation} \label{eq:mlinthelist}
\hat{\xv} = \argmin_{\hat{\xv}_j, j \in \LP 1, \dots, M\RP} \left\Vert \hat{\xv}_j - \yv \right\Vert^2 = \argmax_{\hat{\xv}_j, j \in \LP 1, \dots, M\RP} \sum_{i=0}^{N-1} \hat{x}_{j,i} \cdot y_i,
\end{equation}
where $ \hat{x}_{j,i} \in \LP \pm 1 \RP$, $\hat{\xv}_j$ is the estimated codeword from decoder~$j$ for the received vector $ \yv $ and $\hat{\xv}$ is the final codeword estimate of the ensemble.

\ac{MBBP} is a well-known example for ensemble decoding that uses $ M $ \ac{BP} decoders, each based on a different random parity-check matrix \cite{Huber}. Another instance of ensemble decoding is \ac{BPL} decoding of polar codes, where the stages of the \ac{FFG} are randomly permuted for each constituent decoder \cite{elkelesh2018belief}. 

In this work, we propose \emph{automorphism ensemble decoding} for RM codes with the main idea being to make use of the already existent polar decoders, namely, SC, BP and SCL decoders. 
Furthermore, we use the RM code symmetry in the decoding algorithm itself, as permuting a valid RM codeword with a permutation from the code's automorphism group results in another valid RM codeword.

An abstract view of our proposed decoding algorithm is shown in Fig.~\ref{fig:Block_Diag}.
The input to the decoder is the received noisy codeword $\yv$. 
We randomly sample $M$ different permutations from the RM automorphism group, where each permutation is denoted by $\pi_j$, with $j$ being the decoder index and $j \in \{1,2,\cdots,M\}$.
The $\yv$-vector is interleaved (i.e., permuted) with the $M$ different permutations $\pi_j$ leading to $M$ permuted noisy codewords $\yv'_j$, where $j \in \{1,2,\cdots,M\}$. 
Now we decode every $\yv'_j$-vector using one polar/RM decoder (e.g., BP, SC or SCL) independently and the output from the decoder is the interleaved estimated codeword $\hat{\xv}'_j$.
A de-interleaving phase is applied to all $M$ interleaved estimated codewords $\hat{\xv}'_j$ and, thus, we have the $M$ estimated codewords $\hat{\xv}_j$. Let $ \operatorname{Dec}(\cdot) $ denote the decoding function that maps $ \yv'_j $ to $ \hat{\xv}_j' $, then we can write the interleaved decoding as
\begin{equation}\label{eq:aut_dec}
\hat{\xv}_j = \pi^{-1}_j\left(\operatorname{Dec}\left(\pi_j(\yv)\right)\right).
\end{equation}

Similar to \ac{MBBP} and \ac{BPL} decoding, our proposed decoding algorithm uses the \emph{ML-in-the-list} picking rule according to Eq.~(\ref{eq:mlinthelist}) to choose the most likely codeword from the list to get the final decoder output $\hat{\xv}$. 

As most decoders are linear, their decoding behavior is only dependent on the noise induced by the channel, and not the choice of the transmitted codeword (see Lemma 3 in Appendix~\ref{apx:proof} for a proof of this fact for \ac{SC} decoders). Therefore, decoding using automorphisms according to Eq.~(\ref{eq:aut_dec}) corresponds to permuting the \emph{noise}. It is reasonable to conclude that suboptimal (i.e., not ML) decoders may react differently to noise realizations in different permutations, which is exactly the property that automorphism ensemble decoding seeks to exploit.

Our proposed algorithm can be therefore seen as a natural generalization of the \ac{BPL} decoder \cite{elkelesh2018belief}: We still use $M$ parallel independent decoders; however, we are no longer constrained to BP decoding for the constituent decoder (i.e., we can use \ac{SC}, \ac{BP} or \ac{SCL} decoders as independent constituent decoders).

Furthermore, we use a more general set of permutations. It was shown in \cite{Doan_2018_Permuted_BP} that the stage-shuffling of the \ac{FFG} is equivalent to a bit-interleaving operation while keeping the factor graph unchanged; with the permutations corresponding to the automorphism subgroup $ \Pi(m) $. In contrast, we use permutations from the whole RM code automorphism group $ \operatorname{GA}(m) $ (rather than only $\Pi(m)$, which is used in \ac{BPL} decoding as proposed in \cite{elkelesh2018belief}). 

In the following, \emph{Aut-$M$-Dec} will denote the proposed decoding scheme, with \emph{Dec} being the constituent decoder and $ M $ the ensemble size. If applicable, $ L $ denotes the list size of \ac{SCL} constituent decoders.

It is worth mentioning that the usage of a \ac{BP} decoder as a constituent decoder has some similarities when compared to \emph{automorphism group decoding} of \ac{BCH} and Golay codes for the BEC \cite{HUBER_BCH_Golay} and for the AWGN channel \cite{Dimnik_RRD_HDPC}.
Automorphism group decoding is based on permuting the received sequence exploiting automorphisms of the code while applying an iterative message passing algorithm. 

A similar usage of an SC decoder as constituent decoder was reported in \cite{Moscow_Huawei_RM_paper} and \cite{Moscow_Huawei_Polar_paper}.
However, the used permutations were only constrained to the ones corresponding to stage-shuffling the code's factor graph.
As will be shown later, this constraint degrades the error-rate performance when compared to choosing the permutations from the full automorphism group of the code (i.e., our proposed approach).

\section{Automorphisms of the SC Decoder}\label{sec:auto-SC}
We will now present an interesting fact about \ac{SC} decoding of polar and RM codes using permutations.

\textbf{Theorem 2}:
Permutations $ \pi \in \operatorname{LTA}(m) $ commute with the SC decoding operation of a decreasing monomial code $ \mathcal{C} $, i.e., \begin{equation}\label{eq:thm2}
\operatorname{SC}(\pi(\Lm_{\mathrm{ch}})) = \pi\left(\operatorname{SC}(\Lm_{\mathrm{ch}})\right).
\end{equation}
In other words, it does not matter whether we first permute the received \ac{LLR} vector $ \Lm_{\mathrm{ch}} $ and then decode or decode first and then permute.

\emph{Proof:} The proof is given in Appendix \ref{apx:proof}.

\textbf{Corollary 2.1 (SC decoding with LTA automorphisms)}:
All output candidate codewords from an automorphism-SC decoder are identical. Consequently, there is no gain in using LTA permutations for automorphism-SC decoding (i.e., ensemble decoding with SC as a constituent decoder) compared to a single SC decoder.

\emph{Proof:} The $ j $-th candidate codeword of automorphism-SC decoding with permutation $ \pi_j \in \operatorname{LTA}(m) $ is given by 
\begin{align}
 \hat{\xv}_j = \pi^{-1}_j\left(\operatorname{SC}\left(\pi_j(\Lm_{\mathrm{ch}})\right)\right) &\stackrel{\text{Th.2}}{=} \pi^{-1}_j\left(\pi_j\left(\operatorname{SC}\left(\Lm_{\mathrm{ch}}\right)\right)\right)\nonumber\\ &= \operatorname{SC}\left(\Lm_{\mathrm{ch}}\right),
\end{align}
where Theorem 2 has been applied in the second step.$\qed$

This has far-reaching consequences for polar codes, whose only known automorphisms lie in the $ \operatorname{LTA}(m) $ subgroup. Hence, Automorphism-SC decoding is bound to fail for general polar codes (i.e., leads to the same error-rate performance as a plain SC decoder).

Furthermore, we can combine the results of Theorem 1 and Theorem 2 as follows:

\textbf{Corollary 2.2 (LTA absorption)}:
Automorphism-SC decoding \emph{absorbs} every permutation from $ \operatorname{LTA}(m) $, i.e., for \ac{RM} codes, it is sufficient to use permutations that are a product of a permutation from $ \operatorname{UTA}(m) $ and a permutation from $ \Pi(m) $, rather than the full automorphism group $ \operatorname{GA}(m) $, without any loss in performance.

\emph{Proof:} By Theorem 1, we can factor every permutation $ \pi_j $ from $ \operatorname{GA}(m) $ as $ \pi_j = \pi_{j,L} \circ \pi_{j,U} \circ \pi_{j,P} $. If we apply Automorphism-SC, we have

\begin{align}
\hat{\xv}_j &= (\pi^{-1}_{j,P}\circ\pi^{-1}_{j,U}\circ\pi^{-1}_{j,L})\left(\operatorname{SC}\left((\pi_{j,L} \circ \pi_{j,U} \circ \pi_{j,P})(\Lm_{\mathrm{ch}}) \right) \right) \\ 
&\stackrel{\text{Th.2}}{=} (\pi^{-1}_{j,P}\circ\pi^{-1}_{j,U}\circ\pi^{-1}_{j,L}\circ\pi_{j,L})\left(\operatorname{SC}\left((\pi_{j,U} \circ \pi_{j,P})(\Lm_{\mathrm{ch}}) \right) \right)\\
&= (\pi^{-1}_{j,P}\circ\pi^{-1}_{j,U})\left(\operatorname{SC}\left((\pi_{j,U} \circ \pi_{j,P})(\Lm_{\mathrm{ch}}) \right) \right),
\end{align}
where again Theorem 2 has been applied in the second step.$\qed$

At this point, it should be noted that all this argumentation holds for constituent \ac{SCL} decoders (namely, Automorphism-SCL decoding), as the decoding sequence is identical to the \ac{SC} decoder. 
In addition, as this is both an effect of the factor graph and the decoding procedure, we expect similar (but far less pronounced) behavior also for constituent \ac{BP} decoders, which use the same factor graph. This is confirmed by \ac{BLER} simulations as shown in Section \ref{sec:Results}.

\section{Results}\label{sec:Results}
Regarding practical applications, both error-rate performance and the computational complexity of the decoding scheme have to be considered. We compare the described decoding schemes for the RM(3,7)-code with $ N=128 $ and $ k=64 $ and the RM(4,8)-code with $ N=256 $ and $ k=163 $. In the following, we specify the parameters of the compared decoders for reproducibility:

\begin{itemize}
\item The \textbf{na\"ive \ac{BP}}  and \textbf{\ac{FFG} \ac{BP}} decoders use $ N_\mathrm{it,max} = 200 $ iterations. 
\item \textbf{\ac{MWPC-BP}} utilizes 5\% of the minimum-weight parity-checks and 30 iterations, which is the best-performing configuration reported in \cite{PfisterMWPC}. 
\item \textbf{\ac{MBBP}} operates over $ M=60 $ randomly generated $ \Hm $-matrices with 6 iterations each. 
\item \textbf{Neural-BP} uses all 94488 minimum-weight parity-checks over 6 iterations. 
\item The \textbf{pruned neural-BP} employs on average 3\% of the minimum-weight parity-checks over a total of 6 iterations. We consider the three variants of this decoder as introduced in \cite{buchberger2020pruning}, with tied weights ($ D_1 $), no weights ($ D_2 $) and free weights ($ D_3 $). 
\item Recursive list decoders \textbf{SCL} and its variation using stage-shuffle permutations \textbf{DS} \cite{permuteRM} with list size $ L=32 $.
\item For our proposed \textbf{Aut-BP}, we show results for both $ M=8 $ and $ M=32 $ randomly chosen permutations from the full automorphism group. Here, up to $ N_\mathrm{it,max} = 200 $ iterations are performed with, however, an early stopping condition employed to reduce the average total number of iterations. Furthermore, the \acp{FFG} have been reduced from 1792 to 1334 box-plus and addition operations per full iteration by removing operations with constant results, as presented in Section \ref{sec:ffg}. 
\item Regarding our proposed \ac{SC}-based variant of automorphism ensemble decoding (i.e., \textbf{Aut-SC}), we show results for $ M=8 $ and $ M=32 $. We then compare the effect of the chosen permutations on the error-rate performance of our proposed decoder, both with \ac{BP} and \ac{SC} constituent decoders, using the exemplary ensemble size of $ M=4 $. In particular, we compare permutations from the full automorphism group, upper and lower triangular subgroups $ \operatorname{LTA}(m) $ and $ \operatorname{UTA}(m) $ and permutations corresponding to stage-shuffled factor graphs, i.e., $ \Pi(m) $.
\end{itemize}
\subsection{Error-Rate Performance}
\subsubsection{Comparison with other decoding schemes}

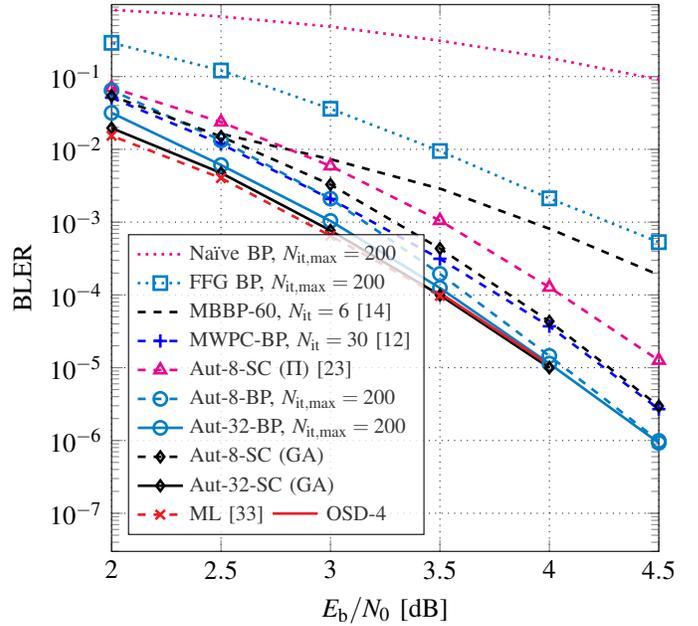
\begin{figure}[t]
	\begin{tikzpicture}
\begin{axis}[
	width=\linewidth,
	height=\linewidth,
	grid style={dotted,anthrazit},
	xmajorgrids,
	yminorticks=true,
	ymajorgrids,
	legend columns=1,
	legend pos=south west,   
	legend cell align={left},
	legend style={fill,fill opacity=0.8},
	xlabel={$E_\mathrm{b}/N_0$ [dB]},
	ylabel={BLER},
	legend image post style={mark indices={}},
	ymode=log,
	mark size=1.5pt,
	xmin=2,
	xmax=4.5,
	ymin=3e-8,
	ymax=9.756e-01
]

\addplot[color=pink,line width = 1pt, dotted, mark size=2.5pt, mark options={solid}]
table[col sep=comma]{
1.00, 9.756e-01
1.50, 9.221e-01
2.00, 8.242e-01
2.50, 6.682e-01
3.00, 4.848e-01
3.50, 3.109e-01
4.00, 1.801e-01
4.50, 9.100e-02
5.00, 3.890e-02
5.50, 1.750e-02
};
\label{plot:rm_bp_naive}
\addlegendentry{\footnotesize Na\"ive BP, $ N_\mathrm{it,max}=200 $ };
\addplot[color=mittelblau,line width = 1pt, dotted,mark=square,mark size=2.5pt, mark options={solid}]
table[col sep=comma]{
1.00, 6.909e-01
1.50, 4.966e-01
2.00, 2.907e-01
2.50, 1.211e-01
3.00, 3.608e-02
3.50, 9.485e-03
4.00, 2.122e-03
4.50, 5.311e-04
5.00, 1.372e-04
};
\label{plot:rm_bp}
\addlegendentry{\footnotesize FFG BP, $ N_\mathrm{it,max}=200 $};

\addplot[color=black,line width = 1pt, dashed,mark size=2.5pt, mark options={solid}]
table[col sep=comma]{
	2.51, 1.612e-02
	3.01, 7.184e-03
	3.51, 2.825e-03
	4.01, 7.922e-04
	4.51, 1.828e-04
	4.71, 1.045e-04
};
\label{plot:rm_mbbp}
\addlegendentry{\footnotesize MBBP-60, $ N_\mathrm{it} = 6 $ \cite{buchberger2020pruning}};

\addplot [color=blue,line width = 1pt, dashed, mark=+, mark size=2.5pt, mark options={solid}]
table[row sep=crcr]{%
	2	0.0512291666666667\\
	2.5	0.011875\\
	3	0.00207945736434109\\
	3.5	0.000313669064748201\\
	4	3.7099161322151e-05\\
	4.5	2.69867196932036e-06\\
};
\addlegendentry{\footnotesize MWPC-BP, $ N_\mathrm{it} = 30 $ \cite{PfisterMWPC}}

\addplot[color=magenta,line width = 1pt, dashed,mark=triangle,mark size=2.5pt, mark options={solid}]
table[col sep=comma]{
1.00, 3.085e-01
1.50, 1.634e-01
2.00, 7.026e-02
2.50, 2.372e-02
3.00, 5.914e-03
3.50, 1.051e-03
4.00, 1.275e-04
4.50, 1.262e-05
};
\label{plot:rm_aut_sc8_pi}
\addlegendentry{\footnotesize Aut-8-SC ($ \Pi $) \cite{Moscow_Huawei_RM_paper}};

\addplot[color=mittelblau,line width = 1pt, dashed,mark=o,mark size=2.5pt, mark options={solid}]
table[col sep=comma]{
	1.00, 3.599e-01
	1.50, 1.766e-01
	2.00, 6.498e-02
	2.50, 1.325e-02
	3.00, 2.103e-03
	3.50, 1.946e-04
	4.00, 1.452e-05
	4.50, 9.887e-07
};
\label{plot:rm_aut_8_bp}
\addlegendentry{\footnotesize Aut-8-BP, $ N_\mathrm{it,max}=200 $};

\addplot[color=mittelblau,line width = 1pt,mark=o, solid,mark size=2.5pt, mark options={solid}]
table[col sep=comma]{
	1.00, 2.776e-01
	1.50, 1.076e-01
	2.00, 3.173e-02
	2.50, 6.082e-03
	3.00, 1.039e-03
	3.50, 1.247e-04
	4.00, 1.139e-05
	4.50, 9.293e-07
};
\label{plot:rm_aut_32_bp}
\addlegendentry{\footnotesize Aut-32-BP, $ N_\mathrm{it,max}=200 $};

\addplot[color=black,line width = 1pt, dashed,mark=diamond,mark size=2pt, mark options={solid}]
table[col sep=comma]{
1.00, 2.792e-01
1.50, 1.520e-01
2.00, 5.427e-02
2.50, 1.500e-02
3.00, 3.283e-03
3.50, 4.340e-04
4.00, 4.338e-05
4.50, 2.966e-06
};
\label{plot:rm_aut_8_sc}
\addlegendentry{\footnotesize Aut-8-SC (GA) };

\addplot[color=black,line width = 1pt, solid,mark=diamond,mark size=2pt, mark options={solid}]
table[col sep=comma]{
1.00, 1.714e-01
1.50, 6.702e-02
2.00, 1.960e-02
2.50, 4.741e-03
3.00, 7.684e-04
3.50, 9.935e-05
4.00, 9.984e-06
};
\label{plot:rm_aut_32_sc}
\addlegendentry{\footnotesize Aut-32-SC (GA) };

\addplot[color=rot,line width = 1pt,mark=x, dashed,mark size=2.5pt, mark options={solid}]
table[col sep=comma]{
	0.00, 5.025e-01
	0.50, 3.205e-01
	1.00, 1.538e-01
	1.50, 5.590e-02
	2.00, 1.538e-02
	2.50, 4.031e-03
	3.00, 6.489e-04
	3.50, 9.700e-05
};
\label{plot:rm_ml}
\addlegendentry{\footnotesize ML \cite{kldatabase} \ref{plot:rm_osd4} OSD-4};

\addplot[color=rot, line width = 1pt, solid, mark size=2.5pt, mark options={solid}]
table[col sep=comma]{
3.50, 1.063e-04
4.00, 1.090e-05
};
\label{plot:rm_osd4}

\end{axis}

\end{tikzpicture}
	\vspace{-0.7cm}
	\caption{\footnotesize BLER comparison between non \ac{SGD}-optimized iterative decoders and our proposed decoding schemes (namely, Aut-BP and Aut-SC) for the RM(3,7)-code over the BI-AWGN channel.}  
	\label{fig:RM_BLER_classic}
\end{figure}

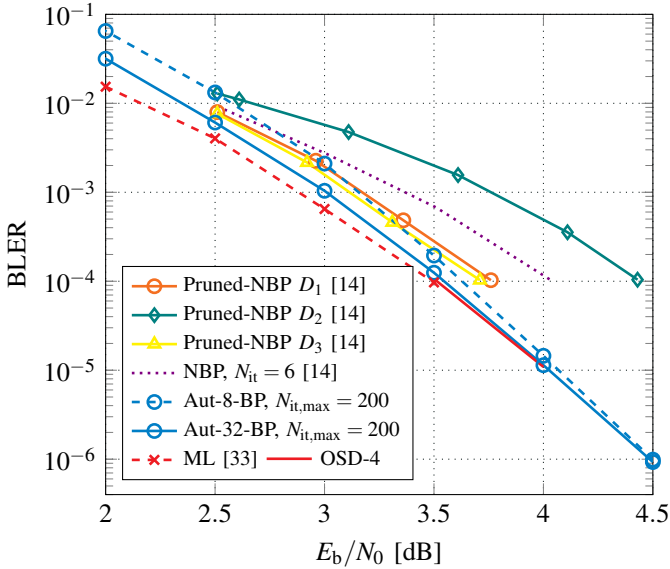
\begin{figure}[t]
	\begin{tikzpicture}
\begin{axis}[
	width=\linewidth,
	height=.9\linewidth,
	grid style={dotted,anthrazit},
	xmajorgrids,
	yminorticks=true,
	ymajorgrids,
	legend columns=1,
	legend pos=south west,   
	legend cell align={left},
	xlabel={$E_\mathrm{b}/N_0$ [dB]},
	ylabel={BLER},
	legend image post style={mark indices={}},
	ymode=log,
	mark size=1.5pt,
	xmin=2,
	xmax=4.5,
	ymin=4e-7,
	ymax=1e-1
]

\addplot[color=orange,line width = 1pt, mark=o, solid,mark size=2.5pt, mark options={solid}]
table[col sep=comma]{
2.51, 7.978e-03
2.96, 2.263e-03
3.36, 4.846e-04
3.76, 1.028e-04
};
\label{plot:rm_d1}
\addlegendentry{\footnotesize Pruned-NBP $ D_1 $ \cite{buchberger2020pruning}};

\addplot[color=teal,line width = 1pt, mark=diamond, solid,mark size=2.5pt, mark options={solid}]
table[col sep=comma]{
2.51, 1.289e-02
2.61, 1.104e-02
3.11, 4.756e-03
3.61, 1.562e-03
4.11, 3.554e-04
4.43, 1.041e-04
};
\label{plot:rm_d2}
\addlegendentry{\footnotesize Pruned-NBP $ D_2 $ \cite{buchberger2020pruning}};

\addplot[color=yellow,line width = 1pt, mark=triangle, solid,mark size=2.5pt, mark options={solid}]
table[col sep=comma]{
2.51, 7.642e-03
2.92, 2.164e-03
3.31, 4.597e-04
3.71, 1.030e-04
};
\label{plot:rm_d3}
\addlegendentry{\footnotesize Pruned-NBP $ D_3 $ \cite{buchberger2020pruning}};

\addplot[color=violet,line width = 1pt, dotted,mark size=2.5pt, mark options={solid}]
table[col sep=comma]{
2.51, 9.236e-03
3.01, 2.698e-03
3.51, 6.749e-04
4.03, 1.044e-04

};
\label{plot:rm_wbp}
\addlegendentry{\footnotesize NBP, $ N_\mathrm{it} = 6 $ \cite{buchberger2020pruning}};

\addplot[color=mittelblau,line width = 1pt, dashed,mark=o,mark size=2.5pt, mark options={solid}]
table[col sep=comma]{
	1.00, 3.599e-01
	1.50, 1.766e-01
	2.00, 6.498e-02
	2.50, 1.325e-02
	3.00, 2.103e-03
	3.50, 1.946e-04
	4.00, 1.452e-05
	4.50, 9.887e-07
};
\label{plot:rm_aut_8_bp}
\addlegendentry{\footnotesize Aut-8-BP, $ N_\mathrm{it,max}=200 $};

\addplot[color=mittelblau,line width = 1pt,mark=o, solid,mark size=2.5pt, mark options={solid}]
table[col sep=comma]{
	1.00, 2.776e-01
	1.50, 1.076e-01
	2.00, 3.173e-02
	2.50, 6.082e-03
	3.00, 1.039e-03
	3.50, 1.247e-04
	4.00, 1.139e-05
	4.50, 9.293e-07
};
\label{plot:rm_aut_32_bp}
\addlegendentry{\footnotesize Aut-32-BP, $ N_\mathrm{it,max}=200 $};

\addplot[color=rot,line width = 1pt,mark=x, dashed,mark size=2.5pt, mark options={solid}]
table[col sep=comma]{
	0.00, 5.025e-01
	0.50, 3.205e-01
	1.00, 1.538e-01
	1.50, 5.590e-02
	2.00, 1.538e-02
	2.50, 4.031e-03
	3.00, 6.489e-04
	3.50, 9.700e-05
};
\label{plot:rm_ml}
\addlegendentry{\footnotesize ML \cite{kldatabase} \ref{plot:rm_osd4} OSD-4};

\addplot[color=rot, line width = 1pt, solid, mark size=2.5pt, mark options={solid}]
table[col sep=comma]{
	3.50, 1.063e-04
	4.00, 1.090e-05
};
\label{plot:rm_osd4}

\end{axis}

\end{tikzpicture}
	\vspace{-0.7cm}
	\caption{\footnotesize BLER comparison between \ac{SGD}-optimized (NN-based) iterative decoders and Aut-BP for the RM(3,7)-code over the BI-AWGN channel. All neural-\ac{BP} decoders use $ N_\mathrm{it} = 6 $ iterations.} 
	\label{fig:RM_BLER_neural}
\end{figure}

In Fig.~\ref{fig:RM_BLER_classic} and Fig. \ref{fig:RM_BLER_neural},  we showcase the error-rate performance of the described decoding schemes for the RM(3,7)-code over the \ac{AWGN} channel using \ac{BPSK} mapping. Furthermore, we show the \ac{ML} performance of the code as provided by \cite{kldatabase}. As no data beyond an \ac{SNR} of 3.5 dB is available, the \ac{ML} performance is estimated using order-4 \ac{OSD}. 

Fig.~\ref{fig:RM_BLER_classic} compares the non-\ac{SGD}-optimized iterative decoders with Aut-BP, Aut-SC and \ac{ML}. One can observe that the na\"ive \ac{BP} decoding suffers from a very poor performance for \ac{RM} codes, compared to \ac{BP} decoding over \ac{FFG}. Moreover, using multiple $ \Hm $-matrices in \ac{MBBP} leads to a significant enhancement in performance. All of the previous methods are outperformed by both Aut-8-BP and \ac{MWPC-BP}, with similar performance. However, in the high \ac{SNR} regime, Aut-8-BP beats \ac{MWPC-BP} by 0.2~dB. Aut-32-BP even closes the gap to \ac{ML} to less than 0.05~dB at a \ac{BLER} of $ 10^{-4} $. Lastly, Aut-32-SC is outperforming all other schemes and is approaching the \ac{ML} performance, while Aut-8-SC is still 0.3 dB away from the \ac{ML} performance. Still, Aut-8-SC sampling from the whole general affine group $ \operatorname{GA}(m) $ outperforms the results from just using stage-shuffle permutations presented in \cite{Moscow_Huawei_RM_paper}.

Fig.~\ref{fig:RM_BLER_neural} compares the \ac{SGD}-optimized (NN-based) decoders with Aut-BP and \ac{ML}. Here, the neural-BP decoder is much closer to the \ac{ML} bound, and the pruned variant with free weights $ D_3 $ outperforms \ac{NBP}, which uses all overcomplete parity-checks. The pruned \ac{NBP} $ D_2 $ decoder without weights suffers from a significant performance degradation. Over the whole SNR range, $ D_1 $ and $ D_3 $ are outperformed by Aut-32-BP. Furthermore, it can be seen that using only $ M=8 $ parallel BP decoders (i.e., Aut-8-BP) results in a small performance degradation of less than 0.2~dB over the whole \ac{SNR} range, offering an attractive trade-off for lower complexity. 

\subsubsection{Comparison to stage-shuffle permutations}

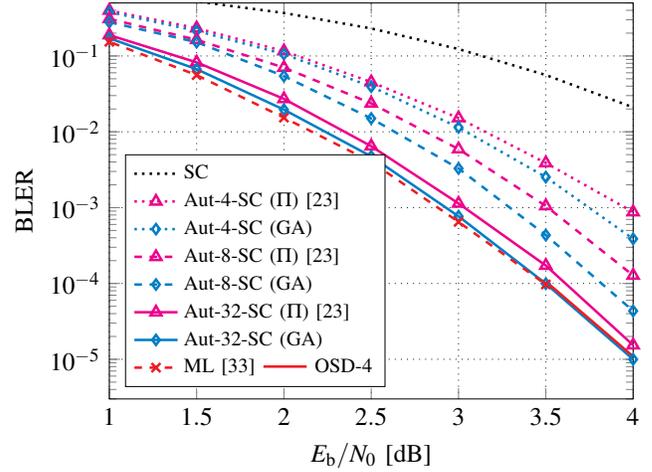
\begin{figure}
	\vspace*{-.3cm}
	\centering
	\resizebox{0.975\columnwidth}{!}{\begin{tikzpicture}
\begin{axis}[
width=\linewidth,
height=.8\linewidth,
grid style={dotted,anthrazit},
xmajorgrids,
yminorticks=true,
ymajorgrids,
legend columns=1,
legend pos=south west,   
legend cell align={left},
xlabel={$E_\mathrm{b}/N_0$ [dB]},
ylabel={BLER},
legend image post style={mark indices={}},
ymode=log,
mark size=1.5pt,
xmin=1,
xmax=4,
ymin=3e-6,
ymax=5e-01
]

\addplot[color=black,line width = 1pt, dotted,mark size=2.5pt, mark options={solid}]
table[col sep=comma]{
1.00, 6.726e-01
1.50, 5.242e-01
2.00, 3.689e-01
2.50, 2.306e-01
3.00, 1.234e-01
3.50, 5.605e-02
4.00, 2.101e-02
4.50, 6.320e-03
5.00, 1.070e-03
5.50, 2.215e-04
6.00, 2.270e-05
};
\label{plot:rm_sc}
\addlegendentry{\footnotesize SC};

\addplot[color=magenta,line width = 1pt, dotted,mark=triangle,mark size=2.5pt, mark options={solid}]
table[col sep=comma]{
	1.00, 3.978e-01
	1.50, 2.327e-01
	2.00, 1.159e-01
	2.50, 4.527e-02
	3.00, 1.519e-02
	3.50, 3.880e-03
	4.00, 8.736e-04
};
\label{plot:rm_aut_sc4_pi}
\addlegendentry{\footnotesize Aut-4-SC ($ \Pi $) \cite{Moscow_Huawei_RM_paper}};

\addplot[color=mittelblau,line width = 1pt, dotted,mark=diamond,mark size=2pt, mark options={solid}]
table[col sep=comma]{
	1.00, 3.793e-01
	1.50, 2.142e-01
	2.00, 1.059e-01
	2.50, 3.947e-02
	3.00, 1.143e-02
	3.50, 2.528e-03
	4.00, 3.875e-04
};
\label{plot:rm_aut_sc4_ga}
\addlegendentry{\footnotesize Aut-4-SC (GA)};

\addplot[color=magenta,line width = 1pt, dashed,mark=triangle,mark size=2.5pt, mark options={solid}]
table[col sep=comma]{
1.00, 3.085e-01
1.50, 1.634e-01
2.00, 7.026e-02
2.50, 2.372e-02
3.00, 5.914e-03
3.50, 1.051e-03
4.00, 1.275e-04
};
\label{plot:rm_aut_sc8_pi}
\addlegendentry{\footnotesize Aut-8-SC ($ \Pi $) \cite{Moscow_Huawei_RM_paper}};

\addplot[color=mittelblau,line width = 1pt, dashed,mark=diamond,mark size=2pt, mark options={solid}]
table[col sep=comma]{
1.00, 2.792e-01
1.50, 1.520e-01
2.00, 5.427e-02
2.50, 1.500e-02
3.00, 3.283e-03
3.50, 4.340e-04
4.00, 4.338e-05
4.50, 2.966e-06
};
\label{plot:rm_aut_sc8_ga}
\addlegendentry{\footnotesize Aut-8-SC (GA)};

\addplot[color=magenta,line width = 1pt, solid,mark=triangle,mark size=2.5pt, mark options={solid}]
table[col sep=comma]{
1.00, 1.887e-01
1.50, 8.237e-02
2.00, 2.717e-02
2.50, 6.458e-03
3.00, 1.135e-03
3.50, 1.722e-04
4.00, 1.536e-05
};
\label{plot:rm_aut_sc32_pi}
\addlegendentry{\footnotesize Aut-32-SC ($ \Pi $) \cite{Moscow_Huawei_RM_paper}};

\addplot[color=mittelblau,line width = 1pt, solid,mark=diamond,mark size=2pt, mark options={solid}]
table[col sep=comma]{
1.00, 1.714e-01
1.50, 6.702e-02
2.00, 1.960e-02
2.50, 4.741e-03
3.00, 7.684e-04
3.50, 9.935e-05
4.00, 9.984e-06
};
\label{plot:rm_aut_sc32_ga}
\addlegendentry{\footnotesize Aut-32-SC (GA)};

\addplot[color=rot,line width = 1pt,mark=x, dashed,mark size=2.5pt, mark options={solid}]
table[col sep=comma]{
	0.00, 5.025e-01
	0.50, 3.205e-01
	1.00, 1.538e-01
	1.50, 5.590e-02
	2.00, 1.538e-02
	2.50, 4.031e-03
	3.00, 6.489e-04
	3.50, 9.700e-05
};
\label{plot:rm_ml}
\addlegendentry{\footnotesize ML \cite{kldatabase} \ref{plot:rm_osd4} OSD-4};

\addplot[color=rot, line width = 1pt, solid, mark size=2.5pt, mark options={solid}]
table[col sep=comma]{
	3.50, 1.063e-04
	4.00, 1.090e-05
};
\label{plot:rm_osd4}

\end{axis}

\end{tikzpicture}}
	\caption{\footnotesize Comparison of using the full automorphism group vs. the stage-shuffle permutations subgroup proposed in \cite{Moscow_Huawei_RM_paper} for the RM(3,7)-code with SC-based constituent decoders; BI-AWGN channel.}
	\label{fig:RM_AutSC}
\end{figure}

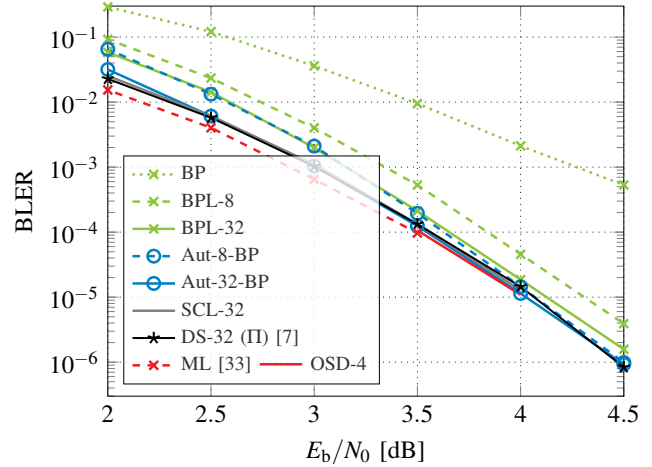
\begin{figure}
	\vspace*{-.3cm}
	\centering
	\resizebox{0.975\columnwidth}{!}{\begin{tikzpicture}
\begin{axis}[
width=\linewidth,
height=.8\linewidth,
grid style={dotted,anthrazit},
xmajorgrids,
yminorticks=true,
ymajorgrids,
legend columns=1,
legend pos=south west,   
legend cell align={left},
legend style={fill,fill opacity=0.8},
xlabel={$E_\mathrm{b}/N_0$ [dB]},
ylabel={BLER},
legend image post style={mark indices={}},
ymode=log,
mark size=1.5pt,
xmin=2,
xmax=4.5,
ymin=3e-07,
ymax=3e-01
]

\addplot[color=apfelgruen,line width = 1pt, dotted,mark=x,mark size=2.5pt, mark options={solid}]
table[col sep=comma]{
	1.00, 6.909e-01
	1.50, 4.966e-01
	2.00, 2.907e-01
	2.50, 1.211e-01
	3.00, 3.608e-02
	3.50, 9.485e-03
	4.00, 2.122e-03
	4.50, 5.311e-04
	5.00, 1.372e-04
};
\label{plot:rm_bp}
\addlegendentry{\footnotesize BP};

\addplot[color=apfelgruen,line width = 1pt, dashed,mark=x,mark size=2.5pt, mark options={solid}]
table[col sep=comma]{
	1.00, 4.807e-01
	1.50, 2.611e-01
	2.00, 9.246e-02
	2.50, 2.340e-02
	3.00, 4.018e-03
	3.50, 5.299e-04
	4.00, 4.559e-05
	4.50, 3.912e-06
};
\label{plot:rm_bpl8}
\addlegendentry{\footnotesize BPL-8};

\addplot[color=apfelgruen,line width = 1pt,mark=x, solid,mark size=2.5pt, mark options={solid}]
table[col sep=comma]{
	1.00, 3.809e-01
	1.50, 1.679e-01
	2.00, 5.829e-02
	2.50, 1.427e-02
	3.00, 1.985e-03
	3.50, 2.119e-04
	4.00, 1.863e-05
	4.50, 1.579e-06
};
\label{plot:rm_bpl32}
\addlegendentry{\footnotesize BPL-32};

\addplot[color=mittelblau,line width = 1pt, dashed,mark=o,mark size=2.5pt, mark options={solid}]
table[col sep=comma]{
1.00, 3.599e-01
1.50, 1.766e-01
2.00, 6.498e-02
2.50, 1.325e-02
3.00, 2.103e-03
3.50, 1.946e-04
4.00, 1.452e-05
4.50, 9.887e-07
};
\label{plot:rm_aut_8_bp}
\addlegendentry{\footnotesize Aut-8-BP };

\addplot[color=mittelblau,line width = 1pt,mark=o, solid,mark size=2.5pt, mark options={solid}]
table[col sep=comma]{
1.00, 2.776e-01
1.50, 1.076e-01
2.00, 3.173e-02
2.50, 6.082e-03
3.00, 1.039e-03
3.50, 1.247e-04
4.00, 1.139e-05
4.50, 9.293e-07
};
\label{plot:rm_aut_32_bp}
\addlegendentry{\footnotesize Aut-32-BP};

\addplot[color=gray,line width = 1pt, solid,mark size=2.5pt, mark options={solid}]
table[col sep=comma, row sep=crcr]{
1.00, 1.880e-01\\
1.50, 7.969e-02\\
2.00, 2.533e-02\\
2.50, 6.105e-03\\
3.00, 1.105e-03\\
3.50, 1.333e-04\\
4.00, 1.275e-05\\
};
\label{plot:rm_scl32}
\addlegendentry{\footnotesize SCL-32};

\addplot[color=black,line width = 0.8pt, solid,mark size=2.5pt,mark=star, mark options={solid}]
table[col sep=comma]{
	1.00, 1.900e-01
	1.50, 7.670e-02
	2.00, 2.276e-02
	2.50, 5.760e-03
	3.00, 1.026e-03
	3.50, 1.324e-04
	4.00, 1.450e-05
	4.50, 8.376e-07
};
\label{plot:rm_ds32}
\addlegendentry{\footnotesize DS-32 ($ \Pi $) \cite{permuteRM} };

\addplot[color=rot,line width = 1pt,mark=x, dashed,mark size=2.5pt, mark options={solid}]
table[col sep=comma]{
	0.00, 5.025e-01
	0.50, 3.205e-01
	1.00, 1.538e-01
	1.50, 5.590e-02
	2.00, 1.538e-02
	2.50, 4.031e-03
	3.00, 6.489e-04
	3.50, 9.700e-05
};
\label{plot:rm_ml}
\addlegendentry{\footnotesize ML \cite{kldatabase} \ref{plot:rm_osd4} OSD-4};

\addplot[color=rot, line width = 1pt, solid, mark size=2.5pt, mark options={solid}]
table[col sep=comma]{
	3.50, 1.063e-04
	4.00, 1.090e-05
};
\label{plot:rm_osd4}

\end{axis}

\end{tikzpicture}}
	\caption{\footnotesize BLER performance of automorphism ensemble decoding of the RM(3,7)-code over the BI-AWGN channel. Comparison of BP, BPL, Aut-BP and recursive list decoders. All iterative decoders use $ N_\mathrm{it,max}=200 $ iterations with early stopping. }
	\label{fig:RM_AutBP}
\end{figure}

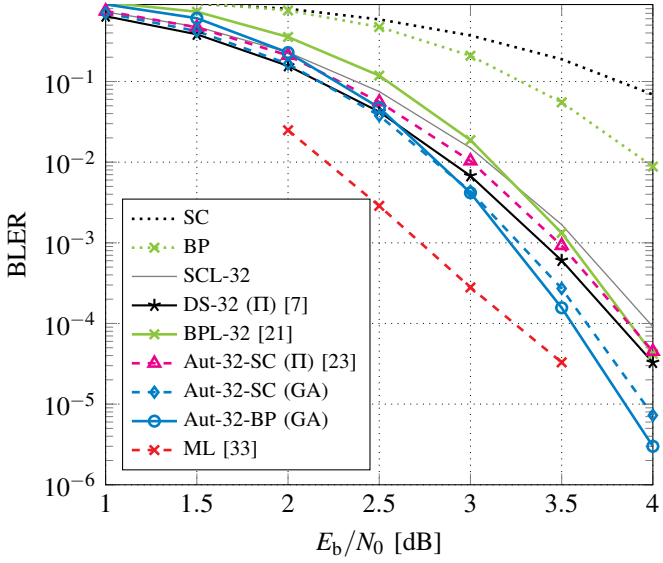
\begin{figure}[t]
		\begin{tikzpicture}
\begin{axis}[
width=\linewidth,
height=.9\linewidth,
grid style={dotted,anthrazit},
xmajorgrids,
yminorticks=true,
ymajorgrids,
legend columns=1,
legend pos=south west,   
legend cell align={left},
xlabel={$E_\mathrm{b}/N_0$ [dB]},
ylabel={BLER},
legend image post style={mark indices={}},
ymode=log,
mark size=1.5pt,
xmin=1,
xmax=4,
ymin=1e-06,
ymax=9e-01
]

\addplot[color=black,line width = 1pt, dotted,mark size=2.5pt, mark options={solid}]
table[col sep=comma]{
	1.00, 9.808e-01
	1.50, 9.207e-01
	2.00, 7.978e-01
	2.50, 5.886e-01
	3.00, 3.725e-01
	3.50, 1.888e-01
	4.00, 6.902e-02
};
\label{plot:rm48_sc}
\addlegendentry{\footnotesize SC};

\addplot[color=apfelgruen,line width = 1pt,mark=x, dotted,mark size=2.5pt, mark options={solid}]
table[col sep=comma]{
1.00, 9.980e-01
1.50, 9.463e-01
2.00, 7.601e-01
2.50, 4.740e-01
3.00, 2.088e-01
3.50, 5.512e-02
4.00, 8.815e-03
};
\label{plot:rm48_bp}
\addlegendentry{\footnotesize BP};

\addplot[color=gray,line width = 0.5pt, solid,mark size=2.5pt, mark options={solid}]
table[col sep=comma]{
	1.00, 7.282e-01
	1.50, 4.784e-01
	2.00, 2.305e-01
	2.50, 7.506e-02
	3.00, 1.500e-02
	3.50, 1.681e-03
	4.00, 9.120e-05
};
\label{plot:rm48_scl32}
\addlegendentry{\footnotesize SCL-32};

\addplot[color=black,line width = 0.8pt, solid,mark size=2.5pt,mark=star, mark options={solid}]
table[col sep=comma]{
1.00, 6.454e-01
1.50, 3.843e-01
2.00, 1.562e-01
2.50, 4.209e-02
3.00, 6.757e-03
3.50, 6.042e-04
4.00, 3.282e-05
};
\label{plot:rm48_ds32}
\addlegendentry{\footnotesize DS-32 ($ \Pi $) \cite{permuteRM} };

\addplot[color=apfelgruen,line width = 1pt, solid, mark=x,mark size=2.5pt, mark options={solid}]
table[col sep=comma]{
1.00, 9.557e-01
1.50, 7.332e-01
2.00, 3.565e-01
2.50, 1.184e-01
3.00, 1.879e-02
3.50, 1.325e-03
4.00, 4.190e-05
};
\label{plot:rm48_bpl32}
\addlegendentry{\footnotesize  BPL-32 \cite{elkelesh2018belief}};

\addplot[color=magenta,line width = 1pt, dashed,mark=triangle,mark size=2.5pt, mark options={solid}]
table[col sep=comma]{
1.00, 7.467e-01
1.50, 4.691e-01
2.00, 2.107e-01
2.50, 5.578e-02
3.00, 1.034e-02
3.50, 9.237e-04
4.00, 4.505e-05
};
\label{plot:rm48_aut_sc32_pi}
\addlegendentry{\footnotesize Aut-32-SC ($ \Pi $) \cite{Moscow_Huawei_RM_paper}};

\addplot[color=mittelblau,line width = 1pt, dashed,mark=diamond,mark size=2pt, mark options={solid}]
table[col sep=comma]{
1.00, 7.066e-01
1.50, 4.212e-01
2.00, 1.655e-01
2.50, 3.780e-02
3.00, 4.344e-03
3.50, 2.716e-04
4.00, 7.261e-06
};
\label{plot:rm48_aut_sc32_ga}
\addlegendentry{\footnotesize Aut-32-SC (GA)};

\addplot[color=mittelblau,line width = 1pt,,mark=o,mark size=2pt, mark options={solid}]
table[col sep=comma]{
	1.00, 9.053e-01
	1.50, 6.090e-01
	2.00, 2.284e-01
	2.50, 4.739e-02
	3.00, 4.169e-03
	3.50, 1.565e-04
	4.00, 3e-06
};
\label{plot:rm48_aut_bp32_ga}
\addlegendentry{\footnotesize Aut-32-BP (GA)};

\addplot[color=rot,line width = 1pt,mark=x, dashed,mark size=2.5pt, mark options={solid}]
table[col sep=comma]{
2.00, 2.488e-02
2.50, 2.873e-03
3.00, 2.797e-04
3.50, 3.294e-05
};
\label{plot:rm48_ml}
\addlegendentry{\footnotesize ML \cite{kldatabase}};

\end{axis}

\end{tikzpicture}
		\vspace{-0.7cm}
		\caption{\footnotesize BLER performance comparison of automorphism ensemble decoding with existing stage permutation-based decoding and \ac{SCL} decoding of the RM(4,8)-code over the BI-AWGN channel. All iterative decoders use $ N_\mathrm{it,max}=200 $ iterations with early stopping. }  
		\label{fig:RM_4_8_BLER}
\end{figure}

We further investigate the gains of sampling from $ \operatorname{GA}(m) $ compared to $ \Pi(m) $ for Aut-SC decoding in Fig. \ref{fig:RM_AutSC} and for Aut-BP decoding in Fig. \ref{fig:RM_AutBP} for the RM(3,7)-code. We can see that for all ensemble sizes $ M $, $ \operatorname{GA}(m) $ consistently outperforms $ \Pi(m) $ by up to 0.3 dB. This confirms the sub-optimality of restricting the automorphisms to a small subgroup. Moreover, Aut-32-BP can even outperform SCL with list size $L=32$ (i.e., SCL-32) and its permutation variant DS-32 in the high SNR regime.

Similar results are obtained for the RM(4,8)-code, depicted in Fig. \ref{fig:RM_4_8_BLER}. The gains of using the full automorphism group are 0.3 and 0.4 dB at a BLER of $ 10^{-3} $ for SC-based and BP-based decoders with ensemble size $ M=32 $, respectively. Here, the proposed automorphism-based decoding schemes outperform SCL-32 consistently and also DS-32 in the high SNR regime.

\subsubsection{SCL-based subdecoders}

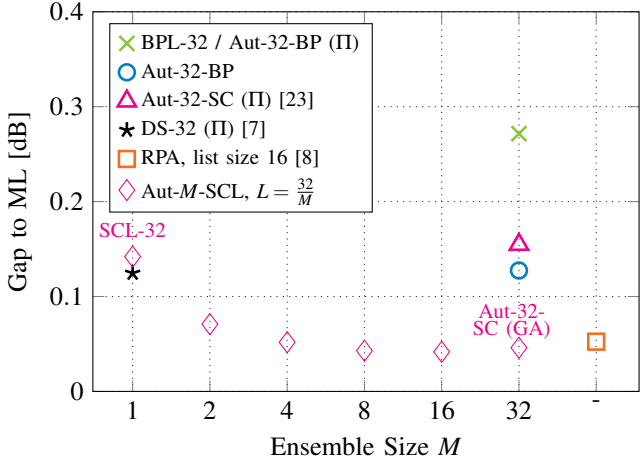
\begin{figure}
	\vspace*{-.3cm}
	\centering
	\resizebox{0.975\columnwidth}{!}{\begin{tikzpicture}
\begin{axis}[
width=\linewidth,
height=.75\linewidth,
grid style={dotted,anthrazit},
xmajorgrids,
yminorticks=true,
ymajorgrids,
legend columns=1,
legend pos=north west,  
xtick={1,2,4,8,16,32,64},
xticklabels={1,2,4,8,16,32,-},
legend cell align={left},
xlabel={Ensemble Size $ M $},
ylabel={Gap to ML [dB]},
legend image post style={mark indices={}},
xmode=log,
mark size=1.5pt,
xmin=.7,
xmax=94,
log basis x=2,
ymin=0,
ymax=0.4
]

\addplot[color=apfelgruen,line width = 1pt, solid,mark=x,mark size=4pt, mark options={solid}, only marks]
table[col sep=comma]{
32,0.2716174079682845
};
\label{plot:rm_bpl32_scat}
\addlegendentry{\footnotesize BPL-32 / Aut-32-BP ($ \Pi $)};

\addplot[color=mittelblau,line width = 1pt, solid,mark=o,mark size=3pt, mark options={solid}, only marks]
table[col sep=comma]{
32,0.12741286543055308
};
\label{plot:rm_aut_bpl32_scat}
\addlegendentry{\footnotesize Aut-32-BP};

\addplot[color=magenta,line width = 1pt, solid,mark=triangle,mark size=4pt, mark options={solid}, only marks]
table[col sep=comma]{
	32,0.15480619514251082
};
\label{plot:rm_aut_sc32_pi_scat}
\addlegendentry{\footnotesize Aut-32-SC ($ \Pi $) \cite{Moscow_Huawei_RM_paper}};

\addplot[color=black,line width = 1pt, solid,mark=star,mark size=3pt, mark options={solid}, only marks]
table[col sep=comma]{
	1,0.1246578
};
\label{plot:rm_dumer_shabunov_scat}
\addlegendentry{\footnotesize DS-32 ($ \Pi $) \cite{permuteRM}};

\addplot[color=orange,line width = 1pt, solid,mark=square,mark size=3pt, mark options={solid}, only marks]
table[col sep=comma]{
	64,0.05242
};
\label{plot:rm_aut_rpa_list16}
\addlegendentry{\footnotesize RPA, list size 16 \cite{RPA_Abbe}};

\addplot[color=magenta,solid,mark=diamond,mark size=4pt, mark options={solid}, only marks]
table[col sep=comma]{
1, 0.14199413933970595
2, 0.07085463520927471
4, 0.05183120045367229
8, 0.042868361249694065
16, 0.041702731333615795
32, 0.04600285190024733
};
\label{plot:aut_scl_scat}
\addlegendentry{\footnotesize Aut-$ M $-SCL, $ L=\frac{32}{M} $};

\node[ magenta] at (1, 0.17) {\footnotesize {SCL-32}};

\node[ magenta] at (30, 0.085) {\footnotesize {Aut-32-}};
\node[ magenta] at (30, 0.065) {\footnotesize {SC (GA)}};

\end{axis}

\end{tikzpicture}}
	\caption{\footnotesize Comparison of automorphism ensemble decoding of the RM(3,7)-code over the BI-AWGN channel at BLER $= 10^{-3} $; 32 codeword candidates; various constituent decoders. For Aut-32-BP and Aut-$ M $-SCL, all automorphisms are randomly sampled from $ \operatorname{GA}(7) $.}
	\label{fig:RM_AutSCL_scatter}
\end{figure}

We study the usage of \ac{SCL}-based constituent decoders. Here, we gain another degree of freedom with the list size $ L $, and denote the decoder by Aut-$ M $-SCL-$ L $. In the comparison we select the parameters $ L $ and $ M $ such that a total of $ L \cdot M = 32 $ codeword candidates is used by all decoder configurations. In Fig.~\ref{fig:RM_AutSCL_scatter} we show the distance to the \ac{ML} bound for each decoder configuration for the RM(3,7)-code. The gap is measured in dB at a \ac{BLER} of $ 10^{-3} $. Note that Aut-1-SCL-32 is a plain \ac{SCL} decoder with $L=32$, and Aut-32-SCL-1 is an Aut-32-SC decoder. We see that, in general, a larger ensemble size $ M $ should be selected rather than a larger list size $ L $ per \ac{SCL} decoder. An optimum is reached for $ L=2 $ and $ M=16 $, i.e., the use of 16 independent SCL-2 decoders, leaving a gap to ML of only 0.04~dB. Again, we see the effect of using the full-automorphism group compared to stage-shuffle permutation also for BP-based constituent decoders, i.e., Aut-$32$-BP vs. BPL-32 \cite{elkelesh2018belief}. Finally, we also compare to \ac{RPA} decoding as proposed in \cite{RPA_Abbe} and observe a slight performance advantage of our scheme. It must be noted, however, that Aut-SC decoding can achieve this performance at only a fraction of the computational complexity of \ac{RPA} decoding.

\subsubsection{Permutation Subgroups}

\begin{figure}[t]
	\centering
	\resizebox{0.975\columnwidth}{!}{\begin{tikzpicture}
\begin{axis}[
width=\linewidth,
height=.75\linewidth,
grid style={dotted,anthrazit},
xmajorgrids,
yminorticks=true,
ymajorgrids,
legend columns=1,
legend pos=south west,   
legend cell align={left},
xlabel={$E_\mathrm{b}/N_0$ [dB]},
ylabel={BLER},
legend image post style={mark indices={}},
ymode=log,
mark size=1.5pt,
xmin=1,
xmax=4,
ymin=1e-04,
ymax=7e-01
]

\addplot[color=black,line width = 1pt, dotted,mark size=2.5pt, mark options={solid}]
table[col sep=comma]{
1.00, 6.726e-01
1.50, 5.242e-01
2.00, 3.689e-01
2.50, 2.306e-01
3.00, 1.234e-01
3.50, 5.605e-02
4.00, 2.101e-02
4.50, 6.320e-03
5.00, 1.070e-03
5.50, 2.215e-04
6.00, 2.270e-05
};
\label{plot:rm_sc}
\addlegendentry{\footnotesize SC};

\addplot[color=apfelgruen,line width = 1pt, dashed,mark=halfsquare left*,mark size=2.5pt, mark options={solid}]
table[col sep=comma]{
1.00, 6.517e-01
1.50, 5.050e-01
2.00, 3.515e-01
2.50, 2.313e-01
3.00, 1.223e-01
3.50, 5.218e-02
4.00, 2.077e-02
};
\label{plot:rm_aut_sc4_lta}
\addlegendentry{\footnotesize Aut-4-SC (LTA)};

\addplot[color=rot,line width = 1pt, dashed,mark=halfsquare right*,mark size=2.5pt, mark options={solid}]
table[col sep=comma]{
1.00, 3.843e-01
1.50, 2.155e-01
2.00, 1.025e-01
2.50, 3.729e-02
3.00, 1.070e-02
3.50, 2.385e-03
4.00, 3.853e-04
};
\label{plot:rm_aut_sc4_uta}
\addlegendentry{\footnotesize Aut-4-SC (UTA)};

\addplot[color=magenta,line width = 1pt, dashed,mark=triangle,mark size=2.5pt, mark options={solid}]
table[col sep=comma]{
	1.00, 3.978e-01
	1.50, 2.327e-01
	2.00, 1.159e-01
	2.50, 4.527e-02
	3.00, 1.519e-02
	3.50, 3.880e-03
	4.00, 8.736e-04
};
\label{plot:rm_aut_sc4_pi}
\addlegendentry{\footnotesize Aut-4-SC ($ \Pi $) \cite{Moscow_Huawei_RM_paper}};

\addplot[color=mittelblau,line width = 1pt, dashed,mark=x,mark size=2pt, mark options={solid}]
table[col sep=comma]{
1.00, 3.793e-01
1.50, 2.142e-01
2.00, 1.059e-01
2.50, 3.947e-02
3.00, 1.143e-02
3.50, 2.528e-03
4.00, 3.875e-04
};
\label{plot:rm_aut_sc4_ga}
\addlegendentry{\footnotesize Aut-4-SC (GA)};

\addplot[color=gray,line width = 0.5pt, solid,mark size=2.5pt, mark options={solid}]
table[col sep=comma]{
1.00, 3.801e-01
1.50, 2.216e-01
2.00, 1.088e-01
2.50, 4.380e-02
3.00, 1.290e-02
3.50, 2.854e-03
4.00, 4.400e-04
4.50, 4.410e-05
};
\label{plot:rm_scl4}
\addlegendentry{\footnotesize SCL-4};

\end{axis}

\end{tikzpicture}}
	\caption{\footnotesize BLER comparison of different automorphism subgroups of the RM(3,7)-code; $M=4$ parallel SC decoders.}
	\label{fig:RM_LTA_UTA_SC}
	\vspace{-0.4cm}
\end{figure}
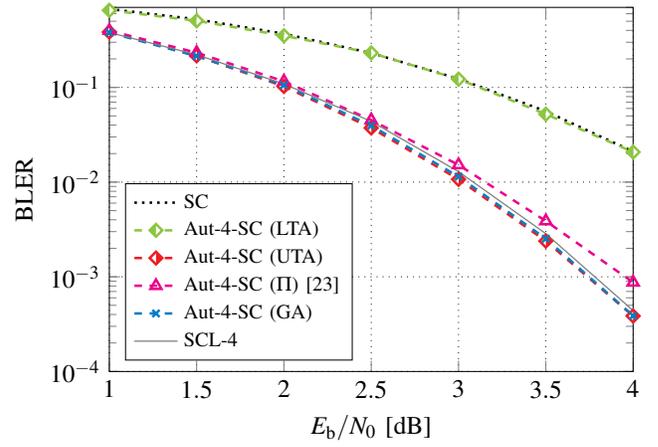

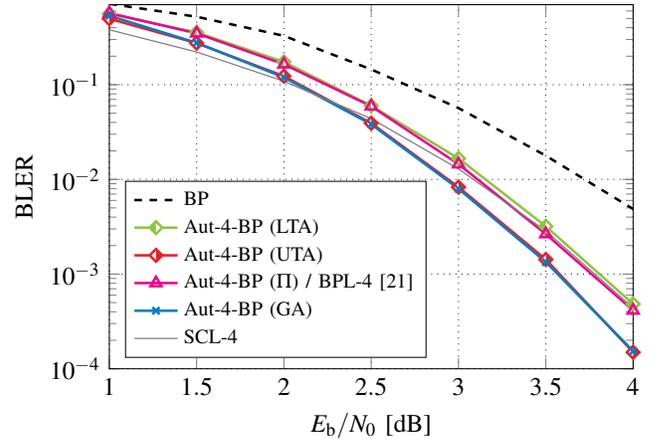
\begin{figure}[t]
	\centering
	\resizebox{0.975\columnwidth}{!}{\begin{tikzpicture}
\begin{axis}[
width=\linewidth,
height=.75\linewidth,
grid style={dotted,anthrazit},
xmajorgrids,
yminorticks=true,
ymajorgrids,
legend columns=1,
legend pos=south west,   
legend cell align={left},
xlabel={$E_\mathrm{b}/N_0$ [dB]},
ylabel={BLER},
legend image post style={mark indices={}},
ymode=log,
mark size=1.5pt,
xmin=1,
xmax=4,
ymin=1e-04,
ymax=7e-01
]

\addplot[color=black,line width = 1pt, dashed,mark size=2.5pt, mark options={solid}]
table[col sep=comma]{
1.00, 7.149e-01
1.50, 5.219e-01
2.00, 3.282e-01
2.50, 1.451e-01
3.00, 5.664e-02
3.50, 1.780e-02
4.00, 4.863e-03
};
\label{plot:rm_bp32}
\addlegendentry{\footnotesize BP};

\addplot[color=apfelgruen,line width = 1pt,mark=halfsquare left*,mark size=2.5pt, mark options={solid}]
table[col sep=comma]{
1.00, 5.610e-01
1.50, 3.567e-01
2.00, 1.741e-01
2.50, 6.006e-02
3.00, 1.665e-02
3.50, 3.195e-03
4.00, 4.831e-04
4.50, 5.360e-05
};
\label{plot:rm_aut_bp4_lta}
\addlegendentry{\footnotesize Aut-4-BP (LTA)};

\addplot[color=rot,line width = 1pt,mark=halfsquare right*,mark size=2.5pt, mark options={solid}]
table[col sep=comma]{
1.00, 4.971e-01
1.50, 2.747e-01
2.00, 1.231e-01
2.50, 3.921e-02
3.00, 8.283e-03
3.50, 1.426e-03
4.00, 1.494e-04
};
\label{plot:rm_aut_bp4_uta}
\addlegendentry{\footnotesize Aut-4-BP (UTA)};

\addplot[color=magenta,line width = 1pt, solid, mark=triangle,mark size=2.5pt, mark options={solid}]
table[col sep=comma]{
	1.00, 5.706e-01
	1.50, 3.487e-01
	2.00, 1.644e-01
	2.50, 5.915e-02
	3.00, 1.449e-02
	3.50, 2.649e-03
	4.00, 4.132e-04
};
\label{plot:rm_bpl4}
\addlegendentry{\footnotesize  Aut-4-BP ($ \Pi $) / BPL-4 \cite{elkelesh2018belief}};

\addplot[color=mittelblau,line width = 1pt,,mark=x,mark size=2pt, mark options={solid}]
table[col sep=comma]{
1.00, 5.351e-01
1.50, 2.769e-01
2.00, 1.193e-01
2.50, 3.817e-02
3.00, 7.939e-03
3.50, 1.339e-03
4.00, 1.538e-04
};
\label{plot:rm_aut_bp4_ga}
\addlegendentry{\footnotesize Aut-4-BP (GA)};

\addplot[color=gray,line width = 0.5pt, solid,mark size=2.5pt, mark options={solid}]
table[col sep=comma]{
	1.00, 3.801e-01
	1.50, 2.216e-01
	2.00, 1.088e-01
	2.50, 4.380e-02
	3.00, 1.290e-02
	3.50, 2.854e-03
	4.00, 4.400e-04
	4.50, 4.410e-05
};
\label{plot:rm_scl4}
\addlegendentry{\footnotesize SCL-4};

\end{axis}

\end{tikzpicture}}
	\caption{\footnotesize BLER comparison of different automorphism subgroups of the RM(3,7)-code; $M=4$ parallel BP decoders; all iterative decoders use $N_{\mathrm{it,max}}~=~32$ iterations with early stopping.}
	\label{fig:RM_LTA_UTA_BP}
	\vspace{-0.4cm}
\end{figure}

Previously, we only focused on the comparison of sampling automorphisms from $ \operatorname{GA}(m) $ and $ \Pi(m) $. In Subsection~\ref{sec:subgroups} we showed that there are more possible automorphism subgroups, namely $ \operatorname{LTA}(m) $ and $ \operatorname{UTA}(m) $. The performance comparison is given in Fig.~\ref{fig:RM_LTA_UTA_SC} for SC-based constituent decoders. While we still use the RM(3,7)-code, we only use a small ensemble size of $ M=4 $ to show the differences more clearly. In order to have general results, and not biased by a bad static selection of permutations, we randomly sample the automorphisms for each simulated codeword. As Corollary 2.1 predicted, the LTA subgroup does not give any gains in ensemble decoding with SC-based decoders, and the \ac{BLER}-curve coincides with plain SC-decoding. More interesting is the fact that UTA automorphisms seem to achieve the same performance as GA, i.e. indicating that components from $ \Pi(m) $ are not required. We further see that Aut-4-SC is slightly outperforming SCL-4.

Fig. \ref{fig:RM_LTA_UTA_BP} shows the results for the same experiment with BP-based constituent decoders, using a maximum of 32 BP iterations each. While not as severe as in the SC-case, LTA automorphisms are again the worst performing subgroup. This can be explained by the similarity of BP and SC decoding, which is performed over the same factor graph. Again, UTA achieves the same performance as GA and both clearly outperform SCL-4 in the high SNR-regime.

\subsubsection{Polar Codes}

\begin{figure}[t]
	\centering
	\resizebox{0.975\columnwidth}{!}{\begin{tikzpicture}[spy using outlines=
{rectangle, magnification=2, connect spies}]
\begin{axis}[
width=\linewidth,
height=.75\linewidth,
grid style={dotted,anthrazit},
xmajorgrids,
yminorticks=true,
ymajorgrids,
legend columns=1,
legend pos=south west,   
legend cell align={left},
legend style={fill,fill opacity=0.8},
xlabel={$E_\mathrm{b}/N_0$ [dB]},
ylabel={BLER},
legend image post style={mark indices={}},
ymode=log,
mark size=1.5pt,
xmin=1,
xmax=4,
ymin=5e-4,
ymax=5e-01
]

\addplot[color=black,line width = 1pt, dotted,mark size=2.5pt, mark options={solid}]
table[col sep=comma]{
1.00, 4.188e-01
1.50, 2.584e-01
2.00, 1.399e-01
2.50, 6.250e-02
3.00, 2.374e-02
3.50, 7.423e-03
4.00, 1.996e-03
};
\label{plot:polar_sc}
\addlegendentry{\footnotesize SC};

\addplot[color=magenta,line width = 1pt, dashed,mark=triangle,mark size=2.5pt, mark options={solid}]
table[col sep=comma]{
1.00, 4.161e-01
1.50, 2.573e-01
2.00, 1.371e-01
2.50, 6.225e-02
3.00, 2.341e-02
3.50, 7.470e-03
4.00, 1.983e-03
};
\label{plot:polar_autsc}
\addlegendentry{\footnotesize Aut-4-SC (random LTA)};

\addplot[color=black,line width = 1pt, dashed,mark size=2.5pt, mark options={solid}]
table[col sep=comma]{
1.00, 4.479e-01
1.50, 2.673e-01
2.00, 1.220e-01
2.50, 4.852e-02
3.00, 1.835e-02
3.50, 5.679e-03
4.00, 1.665e-03
};
\label{plot:polar_bp}
\addlegendentry{\footnotesize BP};

\addplot[color=apfelgruen,line width = 1pt, dashed,mark=x,mark size=2.5pt, mark options={solid}]
table[col sep=comma]{
1.00, 5.271e-01
1.50, 2.904e-01
2.00, 1.425e-01
2.50, 5.757e-02
3.00, 2.109e-02
3.50, 6.838e-03
4.00, 2.001e-03
};
\label{plot:polar_bpl4_random}
\addlegendentry{\footnotesize BPL-4 (random $ \Pi $) \cite{elkelesh2018belief}};

\addplot[color=orange,line width = 1pt,mark=x, solid,mark size=2.5pt, mark options={solid}]
table[col sep=comma]{
1.00, 3.977e-01
1.50, 2.254e-01
2.00, 1.001e-01
2.50, 3.810e-02
3.00, 1.308e-02
3.50, 4.178e-03
4.00, 1.316e-03
};
\label{plot:polar_bpl4_cyclic}
\addlegendentry{\footnotesize BPL-4 (cyclic $ \Pi $) \cite{elkelesh2018belief}};

\addplot[color=mittelblau,line width = 1pt, dashed,mark=o,mark size=2.5pt, mark options={solid}]
table[col sep=comma]{
1.00, 3.618e-01
1.50, 1.937e-01
2.00, 8.812e-02
2.50, 3.189e-02
3.00, 1.090e-02
3.50, 3.810e-03
4.00, 1.165e-03
};
\label{plot:polar_aut_bpl4_random}
\addlegendentry{\footnotesize Aut-4-BP (random LTA)};

\addplot[color=black,line width = 1pt, solid,mark size=2.5pt, mark options={solid}]
table[col sep=comma]{
1.01, 2.190e-01
1.29, 1.540e-01
1.58, 1.020e-01
1.88, 6.410e-02
2.19, 3.750e-02
2.51, 2.190e-02
2.84, 1.180e-02
3.19, 5.880e-03
3.55, 2.790e-03
3.93, 1.240e-03
4.32, 4.630e-04
4.73, 1.600e-04
5.16, 3.320e-05
5.62, 1.510e-05
6.10, 3.030e-06
};
\label{plot:polar_scl32}
\addlegendentry{\footnotesize SCL-32 ($ \approx $ML)};

\addplot[color=gray,line width = 1pt, dashed,mark size=2.5pt, mark options={solid}]
table[col sep=comma]{
	0.00, 5.025e-01
	0.50, 3.205e-01
	1.00, 1.538e-01
	1.50, 5.590e-02
	2.00, 1.538e-02
	2.50, 4.031e-03
	3.00, 6.489e-04
	3.50, 9.700e-05
};
\label{plot:rm_ml}
\addlegendentry{\footnotesize RM(3,7), ML};

\addplot[color=gray,line width = 1pt, dashdotted ,mark size=2.5pt, mark options={solid}]
table[col sep=comma]{
	1.00, 9.878e-02
	1.50, 3.439e-02
	2.00, 7.988e-03
	2.50, 1.274e-03
	3.00, 1.392e-04
	3.50, 1.014e-05
};
\label{plot:polar_crc11_osd}
\addlegendentry{\footnotesize Polar+CRC-11, OSD-4};
\coordinate (spypoint) at (axis cs:3.35,8e-3);
\coordinate (spyviewer) at (axis cs:3.4,1e-1);

\end{axis}

\spy[width=2.2cm,height=1.2cm, fill,fill opacity=0.9] on   (spypoint) in node [fill=white] at (spyviewer);

\end{tikzpicture}}
	\caption{\footnotesize BLER results for the $(N=128,k=64)$ 5G polar code and various constituent decoders. All BP-based decoders use $N_{\mathrm{it,max}}~=~32$ iterations with early stopping. Note that the curves of Aut-4-SC and SC decoding exactly coincide, as predicted by Corollary~2.1.}
	\label{fig:Polar_Autom}
	\vspace{-0.4cm}
\end{figure}
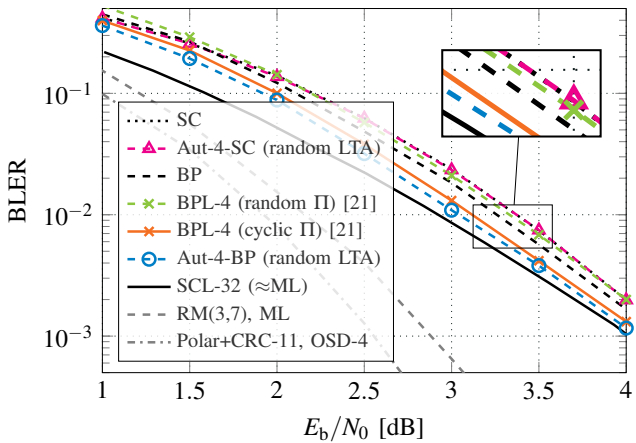

Finally, we also report some results for a polar code, as the proposed decoding schemes conceptually also work with polar codes. The key difference to RM codes is the selection of the information/frozen positions (i.e., $ \mathbb{A} $) and the resulting smaller set of automorphisms, namely $ \operatorname{LTA}(m) $, as pointed out in \cite{bardet_polar_automorphism}. For the sake of reproducibility, we selected the information/frozen-bit positions according to the 5G standard \cite{polar5G2018} and a code length and rate equal to the RM(3,7)-code, i.e., $ k=64 $ and $ N = 128 $. Note that the polar code setup does \emph{not} involve an outer \ac{CRC}. The \ac{BLER} simulation results are presented in Fig.~\ref{fig:Polar_Autom}. We compare Aut-4-SC and Aut-4-BP with SC, SCL, BP and BPL. As expected, Aut-$M$-SC does not work due to Corollary~2.1. However, Aut-4-BP is able to outperform the respective BPL-4 performance and approach the SCL-32 performance, which serves as an estimate of the ML-bound of this polar code. The code design is, however, very suboptimal because it is lacking the outer \ac{CRC} code. Hence, the overall coding/decoding scheme is inferior to \ac{RM} codes under automorphism ensemble decoding. To illustrate this, the ML bound of the RM(3,7)-code and an estimate on the ML bound of the 5G polar code concatenated with the 5G CRC-11 (i.e., Polar+CRC-11) are indicated by the gray lines in Fig.~\ref{fig:Polar_Autom}.

\subsection{Complexity}\label{ssec:Complexity}
\subsubsection{Iterative Decoding}
\begin{figure}[t]
	\resizebox{0.48\textwidth}{!}{\definecolor{mycolor1}{rgb}{0.00000,0.44700,0.74100}%
\definecolor{mycolor2}{rgb}{0.85000,0.32500,0.09800}%
\definecolor{mycolor3}{rgb}{0.69, 0.878,0.902}%
\begin{tikzpicture}

\begin{axis}[%
width=4.3in,
height=2.5in,
at={(0.758in,0.481in)},
scale only axis,
xmin=0,
xmax=12,
ymin=0,
ymax=2300,
ymode=log,
axis background/.style={fill=white},
xlabel style={font=\color{white!15!black}},
xlabel={Decoder},
xticklabels={},
ylabel style={font=\color{white!15!black}},
ylabel={Weighted Complexity [$ \cdot 10^5 $]},
legend columns=6, 
transpose legend,
legend style={at={(0.016,1.45)}, anchor=north west, legend cell align=left, align=left, draw=white!15!black,/tikz/column 2/.style={column sep=5pt}}
]

\addplot[ybar interval, fill=mycolor3, fill opacity=0.6, draw=black, area legend] table[row sep=crcr] {%
	x	y\\
	0.5   263.4\\
	1.5   263.4\\	
};
\addlegendentry{MBBP, $M=60$, $ N_\mathrm{it}=6 $}
\node[rotate=90] at (axis cs: 1,10) {MBBP};

\addplot[ybar interval, fill=blue, fill opacity=0.6, draw=black, area legend] table[row sep=crcr] {%
	x	y\\
	2.5   1853.9\\
	3.5   1853.9\\	
};
\addlegendentry{Neural-BP, $ N_\mathrm{it}=30 $ }
\node[rotate=90] at (axis cs: 3,10) {Neural-BP};

\addlegendimage{empty legend}
\addlegendentry{\hspace{0cm}\textbf{\underline{Pruned-NBP}}}

\addplot[ybar interval, fill=gelb, fill opacity=0.6, draw=black, area legend] table[row sep=crcr] {%
	x	y\\
	4   55.6\\
	5   55.6\\	
};
\addlegendentry{$ D_1 $, $ N_\mathrm{it}=6 $}
\node[rotate=90] at (axis cs: 4.5,10) {$ D_1 $};

\addplot[ybar interval, fill=orange, fill opacity=0.6, draw=black, area legend] table[row sep=crcr] {%
	x	y\\
	5   47.5\\
	6   47.5\\	
};
\addlegendentry{$ D_2 $, $ N_\mathrm{it}=6 $}
\node[rotate=90] at (axis cs: 5.5,10) {$ D_2 $};

\addplot[ybar interval, fill=brown, fill opacity=0.6, draw=black, area legend] table[row sep=crcr] {%
	x	y\\
	6   55.6\\
	7   55.6\\	
};
\addlegendentry{$ D_3 $, $ N_\mathrm{it}=6 $}
\node[rotate=90] at (axis cs: 6.5,10) {$ D_3 $};

\node[] at (axis cs: 5.5,120) {\textbf{Pruned-NBP}};
\addplot[<->,black,line legend,sharp plot,update limits=false,dashed,forget plot] coordinates {(4,100) (7,100)};
\addplot[black,line legend,sharp plot,update limits=false,forget plot] coordinates {(4,55.6) (4,200)};
\addplot[black,line legend,sharp plot,update limits=false,forget plot] coordinates {(7,55.6) (7,200)};

\addplot[ybar interval, fill=aqua, fill opacity=0.6, draw=black, area legend] table[row sep=crcr] {%
	x	y\\
	1.5   463.6\\
	2.5   463.6\\
};
\addlegendentry{MWPC-BP, $ N_\mathrm{it}=30 $}
\node[rotate=90] at (axis cs: 2,10) {MWPC-BP};

\addlegendimage{empty legend}
\addlegendentry{\hspace{0cm}\textbf{\underline{Aut-BP / BPL}}} 

\addplot[ybar interval, fill=apfelgruen, fill opacity=0.6, draw=black, area legend, postaction={pattern=horizontal lines}] table[row sep=crcr] {%
	x	y\\
	7.5   853.8\\
	8.5   853.8\\	
};
\addlegendentry{$M=32$, $ N_\mathrm{it,max}=200 $ (no stopping)}

\addplot[ybar interval, fill=apfelgruen, fill opacity=0.6, draw=black, area legend] table[row sep=crcr] {%
	x	y\\
	8.5   20.6\\
	9.5   20.6\\	
};
\addlegendentry{$M=32$, (with stopping, @ $3.65$ dB)}

\addplot[ybar interval, fill=apfelgruen, fill opacity=0.6, draw=black, area legend, postaction={pattern=vertical lines}] table[row sep=crcr] {%
	x	y\\
	9.5   213.5\\
	10.5   213.5\\	
};
\addlegendentry{$M=8$, $ N_\mathrm{it,max}=200 $ (no stopping)}

\addplot[ybar interval, fill=apfelgruen, fill opacity=0.6, draw=black, area legend, postaction={pattern=dots}] table[row sep=crcr] {%
	x	y\\
	10.5  4.5\\
	11.5  4.5\\	
};
\addlegendentry{$M=8$, (with stopping, @ $3.84$ dB)}

\node[] at (axis cs: 9.5,1450) {\textbf{Aut-BP / BPL}};
\addplot[<->,black,line legend,sharp plot,update limits=false,dashed,forget plot] coordinates {(7.5,1200) (11.5,1200)};
\addplot[black,line legend,sharp plot,update limits=false,forget plot] coordinates {(7.5,853.8) (7.5,2000)};
\addplot[black,line legend,sharp plot,update limits=false,forget plot] coordinates {(11.5,4.5) (11.5,2000)};

\end{axis}
\end{tikzpicture}%
}
	\caption{\footnotesize Complexity comparison of different iterative decoders using basic operations weighted according to Table~\ref{tab:basic_ops} (e.g., weight for multiplication = 3); RM(3,7)-code; BI-AWGN channel.}
	\label{fig:complexity}
\end{figure}
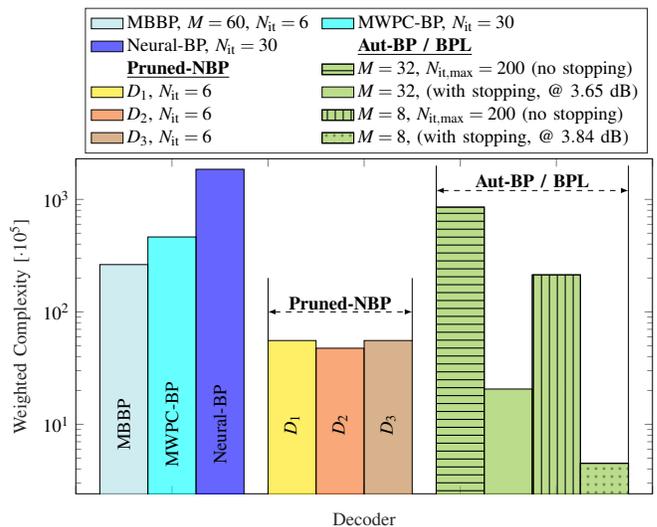

For the RM(3,7)-code, we compare the complexity of the \emph{iterative} decoding algorithms with error-rate performance close to \ac{ML} by counting the number of computing operations required to decode one \ac{RM} codeword. The first column of Table \ref{tab:basic_ops} gives the list of the operations we use.\footnote{Note that these operations differ from the ETSI basic operations, as we are more interested in hardware than in software implementations.} Furthermore, as non-trivial multiplication is significantly more complex than the other considered operations, we introduce a weighting factor for multiplication $ w_\mathrm{mul}=3 $ to make the comparison more fair. For all decoders, we assume that the box-plus operation is implemented as

\begin{align}
L_1 \boxplus L_2 &= \sgn(L_1) \cdot \sgn(L_2) \cdot \min(|L_1|,|L_2|) \nonumber  \\
&\quad+ f_{+}(|L_1+L_2|) - f_{+}(|L_1-L_2|),
\end{align}
where $ f_{+}(|x|) = \log\LB1+\exp(-|x|)\RB$ is a correction term that can be well-approximated by a short \ac{LUT}. Furthermore, \acp{CN} are assumed to be efficiently implemented using the box-minus operator as

\begin{equation}
L_{j\to i} = \bigboxplus_{i' \neq i} L_{i'\to j} = \LB \bigboxplus_{i'} L_{i'\to j} \RB \boxminus L_{i\to j},
\end{equation}
with $ L_1 \boxminus L_2 = \sgn(L_2) \cdot L_1 + f_{-}(|L_1+L_2|) - f_{-}(|L_1-L_2|) $ and $ f_{-}(|x|) = \log(1-\exp(-|x|)) $ which is again implemented as a \ac{LUT} as proposed in \cite{VaryBoxMinus}. The remaining columns of Table \ref{tab:basic_ops} list the number of operations of each type required for the basic building blocks of the described iterative decoding algorithms, namely box-plus evaluations, \ac{CN} and \ac{VN} updates, the ML-in-the-list decision and the stopping condition that is used in \ac{BPL} decoding. Neural-\ac{BP}, the pruned neural-\ac{BP} and \ac{MWPC-BP} decoding use non-trivial multiplications with the corresponding weights before the \ac{VN} evaluations.

Fig. \ref{fig:complexity}
shows the total number of weighted operations to decode one codeword of the RM(3,7)-code. We can see that out of all methods, neural-BP using the full overcomplete $ \Hm $-matrix has the highest complexity. The corresponding pruned decoders $ D_1 $ and $ D_3 $ result in approximately 3\% of that complexity. \ac{MWPC-BP} is computationally more expensive, as it uses more parity-check equations and more iterations are required to achieve a good error-rate performance. It has to be noted however, that we only list the complexity of iterative decoding, not of (adaptively) obtaining the parity-check equations. Hence, the overall complexity of \ac{MWPC-BP} is higher than the presented number. \ac{MBBP} has roughly half the complexity of \ac{MWPC-BP}, while \ac{BPL} without stopping condition has twice the complexity of \ac{MWPC-BP}. However, when a ($ \Gm $-matrix-based) stopping condition is used, the average number of iterations of \ac{BPL} is significantly reduced. Note that even though the ML-in-the-list decision can be only made after all constituent decoders are terminated, terminated  decoders can already start decoding the next received vector (e.g., in a super-scalar implementation). We measure the average required number of iterations until convergence and plot it in Fig. \ref{fig:bpl_iterations} for both $ M=8 $ and $ M=32 $ with respect to the \ac{SNR} of the \ac{AWGN} channel, while $ N_\mathrm{it,max} = 200$. At an \ac{SNR} of 3.65~dB, corresponding to the \ac{BLER} of $ 10^{-4} $, each decoder of the Aut-32-BP ensemble requires an average of 4.55 iterations, making Aut-BP the least complex decoder of the compared algorithms (see Fig.~\ref{fig:complexity}), without losing any error-rate performance. Aut-8-BP requires an \ac{SNR} of 3.84~dB to reach this \ac{BLER} performance, however, reducing the complexity again by a factor of four, using only 3.96 iterations on average.

\begin{table*}[htb]
	\caption{\footnotesize Basic operations and their usage in iterative decoding. \textsuperscript{a}For BP with weights (MWPC-BP, NBP, D1, D3). } \label{tab:basic_ops}
	\centering{\begin{tabular}{l|c|ccccc}
			Operation & Weight & 2-input $ \boxplus $ & CN (deg. $ D $) & VN (deg. $ D $) & ML out of $ M $ & FFG BP Stopping \\
			\hline
			$ \sgn(x)\cdot\sgn(y)$ & 1 & 1 & $ D-1 $ & 0 & 0 & $ m \cdot N/2 + 2N-1 $\\
			$ \sgn(x)\cdot y$ & 1 & 1 & $ 2D-1 $ & 0 & $ MN $ & 0\\
			$ \min(|x|,|y|) $ & 1 & 1 & $ D-1 $ & 0 & 0 & 0\\
			$ \max(x,y) $ & 1 & 0 & 0 & 0 & $ M-1 $ & 0\\
			$ f_{\pm}(|x|) $ (LUT) & 1 & 2 & $ 4D-2 $ & 0 & 0 & 0\\
			$ x+y $,  $ x-y $ & 1 & 4 & $ 8D-4 $ & $ 2D $ & $ M(N-1) $ & 0\\
			$ x\cdot y $ & 3 & 0 & 0 & [$ D+1 $]\textsuperscript{a}  & 0 & 0\\
			\hline
			Weighted total & - & 9 & $ 16D-9 $ & $ 2D $  [$ +3D+3 $]\textsuperscript{a}  & $ 2MN-1 $ & $ m \cdot N/2 + 2N-1 $
	\end{tabular}}
\end{table*}

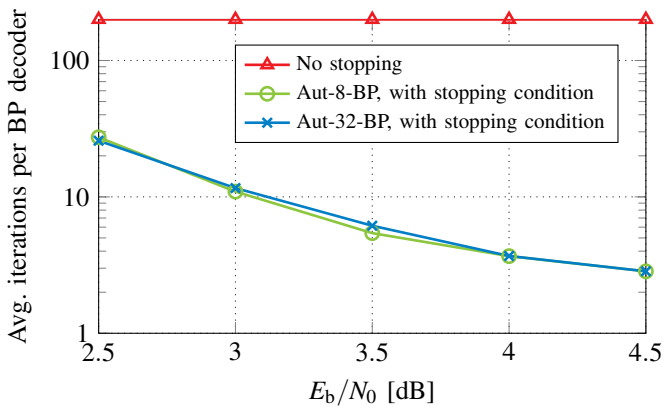
\begin{figure}[t]
	\begin{tikzpicture}
\begin{axis}[
	width=\linewidth,
	height=.65\linewidth,
	grid style={dotted,anthrazit},
	xmajorgrids,
	yminorticks=true,
	ymajorgrids,
	legend style = {
		cells={anchor=west},
		at={(.95,.92)}
	},
	xlabel={$E_\mathrm{b}/N_0$ [dB]},
	ylabel={Avg. iterations per BP decoder},
	ymode=log,
	log ticks with fixed point,
	mark size=1.5pt,
	xmin=2.5,
	xmax=4.5,
	ymin=1,
	ymax=200
]

\addplot[color=rot,line width = 1pt,solid,mark=triangle,mark size=2.5pt,mark options=solid]
table[col sep=comma]{
	2.50, 200
	3.00, 200
	3.50, 200
	4.00, 200
	4.50, 200
};
\label{plot:BPL_200_no};
\addlegendentry{\footnotesize No stopping};

\addplot[color=apfelgruen,line width = 1pt,solid,mark=o,mark size=2.5pt,mark options=solid]
table[col sep=comma]{
	2.50, 27.36
	3.00, 10.93
	3.50, 5.424
	4.00, 3.678
	4.50, 2.8445
};
\label{plot:BPL_8_200_serial};
\addlegendentry{\footnotesize Aut-8-BP, with stopping condition};

\addplot[color=mittelblau,line width = 1pt,solid,mark=x,mark size=2.5pt,mark options=solid]
table[col sep=comma]{
2.50, 2.581e+01
3.00, 1.160e+01
3.50, 6.144e+00
4.00, 3.688e+00
4.50, 2.847e+00
};
\label{plot:BPL_32_200_serial};
\addlegendentry{\footnotesize Aut-32-BP, with stopping condition};

\coordinate (legend) at (axis description cs:.98,.4);

\end{axis}

\end{tikzpicture}
	\vspace{-0.7cm}
	\caption{\footnotesize Average number of iterations required for the constituent BP decoders to converge, with and without early stopping; $ N_\mathrm{it,max}=200 $; RM(3,7)-code.}
	\label{fig:bpl_iterations}
\end{figure}

\subsubsection{Non-Iterative Decoding}
For \ac{SC}-based constituent decoders, the overall complexity is generally much lower than \ac{BP}-based decoding, as one \ac{SC} pass requires roughly the same number of operations as a single \ac{BP} iteration. Instead, the main issue is the latency of the inherently sequential decoding. As there exist many optimized implementations of \ac{SC} and \ac{SCL} decoding dealing with this problem (e.g., \cite{ListDecoder}) and an accurate analysis is rather involved, we do not give numerical complexity analysis here. In general, any optimized version of \ac{SC} and \ac{SCL} may be used in automorphism ensemble decoding, resulting essentially in the same decoding complexity. However, opposed to \ac{SCL}, Aut-SC does not require any sorting operations and therefore is strictly of lower complexity and latency. Similarly, for Aut-SCL decoding, the sizes of the required sorting operations are smaller than in the respective plain \ac{SCL} decoding.
	
\section{Conclusion and Outlook}\label{sec:conc}

In this work, we propose an automorphism-based decoding algorithm for RM codes.
Since polar codes can be seen as a generalization of RM codes, we utilize well-known decoding algorithms of polar codes in the context of RM decoding, namely \ac{SC}, \ac{SCL} and \ac{BP}.
We present near-ML error-rate performance for the RM(3,7)-code (e.g., $0.04$ dB away from the ML bound at BLER of $10^{-3}$).
Furthermore, we report a decoder complexity comparison for the RM(3,7)-code from an operation level perspective.

To the best of our knowledge, our proposed iterative Aut-BP decoders using the RM code automorphism group as permutations are the best iterative decoders reported in literature thus far in terms of error-rate performance when compared to the best previously known iterative decoding schemes for RM codes. Furthermore, the proposed Aut-SC decoders outperform \ac{SCL} decoding both in terms of error-rate and complexity, as no sorting operations are required.

We show that the RM code automorphism group can be divided into three different sub-groups of different performance. 
Based on this classification, we show some theoretical limitations of our proposed decoding algorithm with respect to polar codes.

An interesting open problem is the on-the-fly selection of the best set of $M$ different permutations from the general affine group (i.e., automorphism selection) which are tailored to the specific received noisy codeword $\mathbf{y}$. As previously observed in \cite{CRC_BPL_ISIT20}, this might lead to a significant reduction in the number of automorphisms needed to reach a certain fixed error-rate performance and, thus, reduces the overall complexity of our proposed decoding scheme.

\begin{appendix}

\subsection{Proof of Theorem 2}\label{apx:proof}
Before we can prove Theorem 2, we have to establish a few properties of decreasing monomial codes, SC decoding and LTA permutations.

\textbf{Definition 1 (Upper and Lower Subcode):} Let $ \mathcal{C} $ be a polar code of length $ N=2^m $ which is, with a slight abuse of notation, defined by the logical vector $ \mathbb{A} = \left[a_i\right]$, where $a_i$ denotes whether bit-channel $i$ is a frozen or a non-frozen bit-channel. The \emph{upper and lower subcodes} $ \mathcal{C}_\mathrm{u} $ and $ \mathcal{C}_\mathrm{l} $ of length $ N/2 = 2^{m-1} $ are then given by the first half $ \mathbb{A}_0^{2^{m-1}-1} $ and the second half $ \mathbb{A}_{2^{m-1}}^{2^m-1} $, respectively. The original polar code $ \mathcal{C} $ is formed via the Plotkin construction \cite{ArikanMain,Plotkin} from $ \mathcal{C}_\mathrm{u} $ and $ \mathcal{C}_\mathrm{l} $ as

\begin{equation}\label{eq:plotkin}
\mathcal{C} = \LP \xv = (\xv_\mathrm{u} \oplus \xv_\mathrm{l} | \xv_\mathrm{l}): \quad \xv_u \in \mathcal{C}_\mathrm{u}, \xv_l \in \mathcal{C}_\mathrm{l} \RP.
\end{equation}

\textbf{Lemma 1 (Splitting of Dec. Monomial Codes):} Let $ \mathcal{C} $ be a decreasing monomial code with monomial set $ I $. Then  $ \mathcal{C}_\mathrm{u} $ and $ \mathcal{C}_\mathrm{l} $ are also decreasing monomial codes. Moreover, $ \mathcal{C}_\mathrm{u} $ is completely contained in $ \mathcal{C}_\mathrm{l} $.

\emph{Proof:}
The set of monomials $ I_\mathrm{u} $ of $ \mathcal{C}_\mathrm{u} $ is given by
\begin{equation}\label{eq:info_set_cu}
I_\mathrm{u} = \LP g \div z_{m-1}:\quad g \in I, \,z_{m-1}|g \RP.
\end{equation}
Let $ f,g \in I $ be both divisible by $ z_{m-1} $ and $ f \preccurlyeq g $. Then, by the definition of the partial order, $ f\div z_{m-1} \preccurlyeq g\div z_{m-1}$, as the degrees decrement by one and all other variable indices remain the same. Therefore, $ I_\mathrm{u} $ belongs to a decreasing monomial code.
 
Similarly, the set of monomials $ I_l $ of $ \mathcal{C}_\mathrm{l} $ is given by
\begin{equation}\label{eq:info_set_cl}
I_\mathrm{l} = \LP g:\quad g \in I, \,z_{m-1} \nmid g \RP.
\end{equation}
That this set belongs to a decreasing monomial code follows directly from the definition of the partial order, as the most significant variable index $ m-1 $ is not involved in any monomial.
Lastly, to show that $ \mathcal{C}_\mathrm{u} \subseteq \mathcal{C}_\mathrm{l} $, we only need to show that $ I_\mathrm{u} \subseteq I_\mathrm{l} $. Let $ f\in I_\mathrm{u} $, i.e., $ \exists g\in I $ such that $ g = f\cdot z_{m-1} $. Therefore, $ f \preccurlyeq g $ and $ z_{m-1} \nmid f $ and, thus, $ f \in I_\mathrm{l} $.  $ \qed $

\textbf{Lemma 2 (Pointwise Products):} Let $ \mathcal{C} $ be a decreasing monomial code of length $ N=2^m $ and let $ \mathcal{C}_\mathrm{u} $ and $ \mathcal{C}_\mathrm{l} $ denote its upper and lower subcodes, respectively. Furthermore, let $ \mathcal{C}_\mathrm{RM} $ denote the $ \operatorname{RM}(1,m-1) $ code (i.e., an augmented Hadamard code of length $ 2^{m-1} $). Then the following statement holds:
\begin{equation}\label{eq:pppc}
	\xv_u \in \mathcal{C}_\mathrm{u}, \;\;\; \xv_\mathrm{RM} \in \mathcal{C}_\mathrm{RM} \quad \Rightarrow \quad \xv_\mathrm{u} \odot \xv_\mathrm{RM} \in \mathcal{C}_\mathrm{l},
\end{equation}
where `$ \odot $' denotes a pointwise (i.e., component-wise) multiplication of two vectors.

\emph{Proof:}
By definition, the monomial set $ I_\mathrm{RM} $ of the $ \operatorname{RM}(1,m-1) $ code is
\begin{equation}\label{eq:info_set_rm}
I_\mathrm{RM}  = \LP 1 \RP \cup \LP z_j: \quad 0\le j < m-1 \RP.
\end{equation}

The pointwise product of interest is given by 
\begin{align}\label{eq:poly_eval_pp}
\xv_\mathrm{u} \odot \xv_\mathrm{RM} &= \operatorname{eval}\left(u_\mathrm{u}(\zv)\cdot u_\mathrm{RM}(\zv)\right) \nonumber\\
&= \operatorname{eval}\left( (\uv_\mathrm{u} \cdot \gv_u^T(\zv))\cdot (\uv_\mathrm{RM} \cdot \gv_\mathrm{RM}^T(\zv))\right) \nonumber\\
&= \sum_{i=0}^{|I_\mathrm{u}|-1}\sum_{j=0}^{|I_\mathrm{RM}|-1} \operatorname{eval}\left(u_{\mathrm{u},i} \cdot u_{\mathrm{RM},j} \cdot g_{\mathrm{u},i}(\zv) \cdot g_{\mathrm{RM},j}(\zv) \right)\nonumber\\
&= \operatorname{eval}\left( \sum_{k=0}^{K-1} \tilde{u}_k \cdot \tilde{g}_k(\zv) \right),
\end{align}
with some $ \tilde{u}_k \in \FF_2 $ and $ \tilde{g}_k(\zv) $ monomials from all possible products 
\begin{align}\label{eq:pppc_proof}
I_\mathrm{prod} &= I_\mathrm{u} \odot I_\mathrm{RM} \nonumber \\
&\triangleq \LP f\cdot h: \quad f\in I_\mathrm{u}, h \in I_\mathrm{RM} \RP \nonumber\\
&=\LP f\cdot z_j: \quad f\in I_\mathrm{u}, 0\le j<m-1 \RP \cup \LP f\cdot 1: \quad f\in I_\mathrm{u} \RP \nonumber\\
&=\LP g'=\underbrace{g\div z_{m-1}\cdot z_j}_{\preccurlyeq g \Rightarrow g' \in I}: \, g\in I, z_{m-1}|g, 0\le j<m-1 \RP \cup I_\mathrm{u} \nonumber\\
&\subseteq I_\mathrm{l} \cup I_\mathrm{u} \subseteq I_\mathrm{l},
\end{align}
where we used that $ z_{m-1} \nmid g' $ and, thus, $ g' \in I_\mathrm{l} $ and the fact that $ I_\mathrm{u} \subseteq I_\mathrm{l} $, as established in Lemma 1. Note that any square terms $ z_j^2 $ in the product are implicitly replaced by $ z_j $, since they evaluate to the same expression over $ \FF_2 $. As any $ \xv_\mathrm{u} \odot \xv_\mathrm{RM} $ is generated by the encoding rule of the polar code with monomial set $ I_\mathrm{prod} $, it is also a codeword of $ \mathcal{C} _\mathrm{l} $. $ \qed $

Note that the inverse is not necessarily true, i.e., there might be a codeword in $ \mathcal{C}_\mathrm{l} $ that cannot be written as a pointwise product $ \xv_\mathrm{u} \odot \xv_\mathrm{RM} $.

\textbf{Lemma 3 (Decoder Linearity):} For any received LLR vector $ \Lm $ and any codeword $ \xv \in \mathcal{C} $, we have the property
\begin{equation}
	\uv' = \operatorname{SC}\left(\Lm \odot (-1)^{\xv}\right) = \operatorname{SC}\left(\Lm\right)\oplus \xv,
\end{equation}
where $ \uv' $ denotes the codeword estimate of \ac{SC} decoding. In other words, if we flip the signs of the components of $ \Lm $ according to some codeword $ \xv $, the \ac{SC} codeword estimate will shift exactly by $ \xv $, i.e., the \ac{SC} decoder does not favor any codeword over another.

\emph{Proof:} We will prove this Lemma via induction over the the dimension $ m $ of the code. The base case for $ m=0 $ is straightforward:
\begin{equation}\label{eq:pf_lemma_0_base}
u' = \operatorname{SC}\left((-1)^x L\right) = \begin{cases}
0 = u \oplus x & \text{frozen}\\
\operatorname{HD}\left((-1)^x L\right)=u\oplus x &\text{non-frozen}
\end{cases}
\end{equation}

In the frozen case we used the fact that the only valid codeword is $ x=0 $.

For the induction step $ m \mapsto m+1 $ we use $ \mathcal{C}_\mathrm{u} $ and $ \mathcal{C}_l $ according to Def. 1. By applying Eq. (\ref{eq:first_child_llr}), we find that 
\begin{align}
L_{i,m}' &= \left( (-1)^{x_i} \cdot L_{i,m+1} \right) \boxplus \left( (-1)^{x_j} \cdot L_{j,m+1} \right)\nonumber\\
	&= \left(L_{i,m+1}\boxplus L_{j,m+1}\right)\cdot (-1)^{x_i \oplus x_j}\nonumber\\
	&= L_{i,m}\cdot (-1)^{x_i \oplus x_j},
\end{align}
where we define $ j=i+2^m $ for easier readability. As $ [x_i]~=~\xv_\mathrm{u}~\oplus~\xv_\mathrm{l} $ and $ [x_j] = \xv_\mathrm{l} $, we know that $ [x_i~\oplus~x_j]~=~\xv_\mathrm{u}~\oplus~\xv_\mathrm{l}~\oplus~ \xv_\mathrm{l}~=~\xv_\mathrm{u}~\in~\mathcal{C}_\mathrm{u} $ is in fact a codeword and we can use the induction assumption for the upper branch to claim that
\begin{equation}
u_{i,m}' = u_{i,m} \oplus x_i \oplus x_j
\end{equation}

Similarly, the lower branch equates to
\begin{align}
L_{j,m}' &= (-1)^{u_{i,m} \oplus x_i \oplus x_j} \cdot (-1)^{x_i} \cdot L_{i,m+1} + (-1)^{x_j} \cdot L_{j,m+1} \nonumber\\
&=  (-1)^{u_{i,m} \oplus x_i \oplus x_j \oplus x_i} \cdot L_{i,m+1}  +  (-1)^{x_j} \cdot L_{j,m+1} \nonumber\\
&= L_{j,m}\cdot (-1)^{x_j}.
\end{align}
Obviously, $ [x_j]\in \mathcal{C}_\mathrm{l} $, so we can use the induction assumption for the lower branch to claim that
\begin{equation}
u_{j,m}' = u_{j,m} \oplus x_j.
\end{equation}

The hard decision propagates right as 
\begin{align}
u_{i,m+1}' &= u_{i,m} \oplus x_i \oplus x_j \oplus u_{j,m} \oplus x_j = u_{i,m+1} \oplus x_i,\\
u_{j,m+1}' &= u_{j,m} \oplus x_j = u_{j,m+1} \oplus x_j,
\end{align}
which completes the proof. $ \qed $

\textbf{Lemma 4 (Separation Conservation):} Let $ \pi \in \operatorname{LTA}(m) $ and let $ i_1 $ and $ i_2 $ be two indices that only differ in their \ac{MSB}, i.e., $ |i_1-i_2|=2^{m-1} $. Then, also their images $ i_1'=\pi(i_1) $ and $ i_2'=\pi(i_2) $ differ only in their \ac{MSB}, i.e., $ |\pi(i_1)-\pi(i_2)|=2^{m-1} $.

\emph{Proof:} Let $ \zv $ be the binary representation of $ i_1 $, i.e., $ i_1 =\sum_{j=0}^{m-1} z_j\cdot 2^j $, and $ i_2 $ with binary expansion $ \tilde{\zv} $ such that $ |i_1-i_2|=2^{m-1} $. Without loss of generality, assume $ i_1 < i_2 $, which implies $ i_1 < 2^{m-1} $, i.e., $ z_{m-1}=0 $ and $ i_2 = i_1+2^{m-1} $. Furthermore, let $ \zv' = \Am \zv + \bv \mod 2$ be the binary expansion of the permuted index $ i_1'=\pi(i_1) $. Its bits are given as
\begin{align}
z_j' &= \sum_{k=0}^{m-1}A_{j,k}\cdot z_k+b_j \mod 2 \nonumber\\
&=\sum_{k=0}^{m-2}A_{j,k}\cdot z_k+A_{j,m-1}\cdot\underbrace{z_{m-1}}_{=0} +b_j \mod 2\nonumber\\
&=\sum_{k=0}^{m-2}A_{j,k}\cdot z_k+b_j\mod 2
\end{align}

Then $ \tilde{\zv} $ is
\begin{equation}\label{key}
\tilde{z}_j = \begin{cases}
z_j& j \ne m-1\\
\neg z_j & j= m-1
\end{cases},
\end{equation}
where `$ \neg $' denotes the boolean negation operator.
The image of $ i_2 $ can be computed as
\begin{align}
\tilde{z}_j' &= \sum_{k=0}^{m-1}A_{j,k}\cdot \tilde{z}_k+b_j \mod 2 \nonumber\\
&=\sum_{k=0}^{m-2}A_{j,k}\cdot z_k+ A_{j,m-1}\cdot\underbrace{\tilde{z}_{m-1}}_{=1} +b_j \mod 2\nonumber\\
&=\sum_{k=0}^{m-2}A_{j,k}\cdot z_k+\underbrace{A_{j,m-1}}_{=1 \text{ only for } j=m-1}+b_j \mod 2\nonumber\\
&=\begin{cases}
z_j'& j \ne m-1\\
\neg z_j' & j= m-1
\end{cases},
\end{align}
where we used that $ \Am $ is lower triangular in the last step. $ \qed $

\begin{figure}[t]
	\centering{\begin{tikzpicture}[yscale=1.5]
	\tikzstyle{arrow} = [dspconn,line width = 0.25mm];
	\tikzstyle{normalline} = [line width = 0.5mm];
	\tikzstyle{block} = [rectangle, draw, minimum size=1.5cm, fill=white!90!gray],

	\node[] (p1) at (-3.7, -2) {}; 
	\node[] (p4) at (.7, 2) {};

	\node[block] (SC1) at (-2.75, 1) {$ \operatorname{SC}_{m-1} $};
	\node[block] (SC2) at (-2.75, -1) {$ \operatorname{SC}_{m-1} $};
	
	\node[dspnodefull,minimum size=1.5mm] (temp) at (0, -1) {};
	\node[dspadder](xor) at (0, 1) {};
	\node[] (x1) at (2, 1) {};
	\node[] (x2) at (2, -1) {};
	\draw[normalline](SC1)--(xor)--(x1);
	\draw[normalline](SC2)--(temp)--(x2);
	\draw[normalline](xor)--(temp);

	\draw[arrow]  (-.3,1.2) --node[above] {$L'_{i,m-1}$}(-1.8,1.2);
	\draw[arrow]  (-1.8,.8) --node[below] {$u'_{i,m-1}$}(-.3,0.8);
	
	\draw[arrow]  (-.3,-0.8) --node[above] {$L'_{i+2^{m-1},m-1}$}(-1.8,-0.8);
	\draw[arrow]  (-1.8,-1.2)--node[below] {$u'_{i+2^{m-1},m-1}$}(-.3,-1.2);
	
	\draw[arrow]  (1.8,1.2) --node[above, xshift=2ex] {$L'_{i,m}=L_{\pi(i),m}$}(.3,1.2);
	\draw[arrow]  (.3,.8)   --node[below] {$u'_{i,m}$}(1.8,0.8);
	
	\draw[arrow]  (1.8,-0.8) --node[above, xshift=7ex] {$L'_{i+2^{m-1},m}=L_{\pi(i+2^{m-1}),m}$}(.3,-0.8);
	\draw[arrow]  (.3,-1.2)  --node[below] {$u'_{i+2^{m-1},m}$}(1.8,-1.2);
	
	\draw[<->]  (1,.4) to [out=-35,in=35] (1,-.4);
	\node[scale=1] (text) at (2, 0) {if $ \xi_i = 1 $};
						
\end{tikzpicture}}
	\caption{\footnotesize{Permuted successive cancellation (SC) decoding.}}
	\label{fig:sc_proof}
\end{figure}
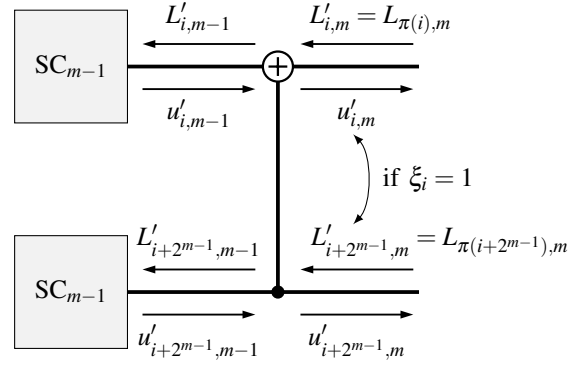
\textbf{Main proof of Theorem 2:} We are now ready to prove Theorem 2 via a form of structural induction over the \ac{SC} decoding tree. Note that this proof therefore assumes the same processing order for the induction as the decoding rules of the \ac{SC} decoder. We regard the right-most stage (i.e., $ s=m $) of the \ac{SC} decoder. Let $ i < 2^{m-1} $ be the index of a processing element of \acp{LLR} in the permuted graph and $ i'=\pi(i) $ its image under the LTA permutation $ \pi $, i.e., the corresponding unpermuted index. From Lemma 4, we know that $ |\pi(i)-\pi(i+2^{m-1})| = 2^{m-1} $. Let $ j $ be always the upper (i.e., numerically smaller) permuted index of the processing element, which can be expressed as
\begin{equation}
j = \min\left(\pi(i), \pi(i+2^{s-1})\right) = \pi(i) \mod 2^{m-1}
\end{equation}

Furthermore, let the binary variable $ \xi_i $ denote, whether $ \pi $ swaps the indices $ i $ and $ i+2^{s-1} $, i.e. 
\begin{equation}
\xi_i = \begin{cases}
0&\pi_s(i) < \pi_s(i+2^{s-1})\\
1&\pi_s(i) > \pi_s(i+2^{s-1})\\
\end{cases}.
\end{equation}

The variables $ \xi_i $ and $ j $ therefore correspond to the \ac{MSB} of $ \pi(i) $ and the rest (i.e., all bits but the \ac{MSB}) of $ \pi(i) $, respectively.

The inputs to the regarded processing element are given by the permuted \acp{LLR}
\begin{align}
L'_{i,m}&=L_{\pi(i),m} \\
L'_{i+2^{m-1},m}&=L_{\pi(i+2^{m-1}),m},
\end{align}
as it is assumed in the conditions for Theorem 2. Fig.~\ref{fig:sc_proof} shows a summary of the definitions in the block diagram of the permuted \ac{SC} decoding stage. 

The unpermuted upper branch is computed according to Eq.~(\ref{eq:first_child_llr}) as
\begin{equation}\label{eq:unp_upper}
L_{j,m-1} = L_{j,m} \boxplus L_{j+2^{m-1},m}.
\end{equation}

In the permuted case, we have
\begin{align}\label{eq:p_upper}
L'_{i,m-1} &= L'_{i,m} \boxplus L'_{i+2^{m-1},m} \nonumber\\
&= L_{\pi(i),m} \boxplus L_{\pi(i+2^{m-1}),m} \nonumber \\
&= \begin{cases}
L_{j,m} \boxplus L_{j+2^{m-1},m} & \text{for } \xi_i=0\\
L_{j+2^{m-1},m} \boxplus L_{j,m} & \text{for } \xi_i=1\\
\end{cases}\nonumber\\
&= L_{j,m-1}\\
&= L_{\tilde{\pi}(i),m-1},
\end{align}
where Lemma 4 was used in the third equality and the fact that `$ \boxplus $' is commutative in the last step. From Eq. (\ref{eq:p_upper}) follows that the inputs to the upper sub-\ac{SC} decoder are unchanged in value, but permuted according to 
\begin{equation}
\tilde{\pi}(i) \triangleq j = \pi(i) \mod 2^{m-1},
\end{equation}
for $ 0\le i < 2^{m-1} $. This permutation $ \tilde{\pi} $ belongs to the affine transform $ \tilde{\zv}' = \tilde{\Am} \tilde{\zv} + \tilde{\bv} \mod 2 $ with $ \tilde{\Am}= \Am_{0,0}^{m-2,m-2} $ and $ \tilde{\bv}= \bv_{0}^{m-2} $, i.e., $ \tilde{\pi} \in \operatorname{LTA}(m-1) $. From Lemma 1 we know that the sub-\ac{SC} decoders on the left also belong to decreasing monomial codes. As a consequence, we can recursively apply Eq. (\ref{eq:p_upper}) until $ m=1 $. Here, we have $ \tilde{\pi}(i)=i $ and $ L'_{0,0}=L_{0,0} $. Therefore
\begin{equation}
u'_{0,0} = u_{0,0},
\end{equation}
which serves as the base case for the left-to-right induction step. Thus, we can now assume that there is an $ m $ for which 
\begin{equation}
u'_{i,m-1} = u_{j,m-1}
\end{equation}
holds and we can use this result for the lower branch. The unpermuted lower branch is computed according to Eq. (\ref{eq:second_child_llr}) as

\begin{equation}\label{eq:unp_lower}
L_{j+2^{m-1},m-1} = (-1)^{u_{j,m-1}} \cdot L_{j,m} + L_{j+2^{m-1},m}.
\end{equation}

In the permuted case, we have
\begin{align}\label{eq:p_lower}
L'_{i+2^{m-1},m-1} &= (-1)^{u'_{i,m-1}} \cdot L'_{i,m} + L'_{i+2^{m-1},m} \nonumber\\
&= \begin{cases}
(-1)^{u_{j,m-1}} \cdot L_{j,m} + L_{j+2^{m-1},m} & \text{for } \xi_i=0\\
(-1)^{u_{j,m-1}} \cdot L_{j+2^{m-1},m}+L_{j,m} & \text{for } \xi_i=1\\
\end{cases}\nonumber\\
&=(-1)^{u_{j,m-1} \cdot \xi_i} \cdot L_{j+2^{m-1},m-1} \\
&=(-1)^{u_{\tilde{\pi}(i),m-1} \cdot \xi_i} \cdot L_{\tilde{\pi}(i)+2^{m-1},m-1}.
\end{align}

This means we have again the same \acp{LLR} values, however (again) permuted by $ \tilde{\pi} $ and flipped in their signs according to $ u_{\tilde{\pi}(i),m-1}\cdot \xi_i $. By expanding the definition of $ \xi_i $, with $ \zv $ denoting the binary expansion of $ i $, we find that
\begin{align}\label{eq:xi_is_cw}
\xi_i = z'_{m-1} &= \Am_{m-1} \cdot \zv + b_{m-1} \mod 2 \nonumber \\
&= q_0 z_0 \oplus \cdots \oplus q_{m-2} z_{m-2} \oplus q_{m-1} \underbrace{z_{m-1}}_{=0} \oplus b_{m-1}, 
\end{align}
for $ i < 2^{m-1} $, which is exactly the definition of an $ \operatorname{RM}(1,m-1) $ codeword, i.e., $ \bm{\xi} = \left[\xi_i \right] \in \operatorname{RM}(1,m-1) $. As $ \uv_{m-1} = \left[u_{\tilde{\pi}(i),m-1} \right] \in \mathcal{C}_u $ because it is the output of an \ac{SC} decoder, we know from Lemma 2, that 
\begin{equation}\label{eq:lower_cw}
\vv_{m-1} = \uv_{m-1}\odot \bm{\xi} \in \mathcal{C}_l.
\end{equation}
Via Lemma 3, we can thus safely ignore the sign-flip due to the pointwise product $ \uv_{m-1}\odot \bm{\xi} $, if we compensate for it on the output:
\begin{align}\label{eq:p_lower_hard}
\left[u'_{i+2^{m-1},m-1}\right] &= \operatorname{SC}\left( \left[ (-1)^{u_{\tilde{\pi}(i),m-1}\cdot \xi_i} \cdot L_{\tilde{\pi}(i)+2^{m-1},m-1} \right]\right) \nonumber\\
&= \operatorname{SC}\left( \left[ L_{\tilde{\pi}(i)+2^{m-1},m-1} \right]\right) \oplus \left[ u_{\tilde{\pi}(i),m-1}\cdot \xi_i  \right] 
\end{align}
With the induction assumption, we have
\begin{equation}\label{eq:lower_lemma_3}
\operatorname{SC}\left( \left[ L_{\tilde{\pi}(i)+2^{m-1},m-1} \right]\right) = \left[ u_{\tilde{\pi}(i)+2^{m-1},m-1} \right]
\end{equation}
and thus
\begin{equation}\label{eq:p_lower_hard_2}
u'_{i+2^{m-1},m-1} = u_{j+2^{m-1},m-1} \oplus \left( u_{j,m-1} \cdot \xi_i \right).
\end{equation}
Finally, we can compute the right-propagating hard decision and decoder output as 
\begin{align}\label{eq:p_final1}
u'_{i,m} &= u'_{i, m-1} \oplus u'_{i+2^{m-1},m-1} \nonumber\\
&= u_{j, m-1} \oplus u_{j+2^{m-1},m-1} \oplus \left( u_{j,m-1} \cdot \xi_i \right) \nonumber\\
&= \begin{cases}
u_{j,m-1} \oplus u_{j+2^{m-1},m-1} = u_{j,m}& \text{for } \xi_i=0\\
u_{j+2^{m-1},m-1} = u_{j+2^{m-1},m}& \text{for } \xi_i=1\\
\end{cases}\nonumber \\
&= u_{\pi(i),m}
\end{align}
and
\begin{align}\label{eq:p_final2}
u'_{i+2^{m-1},m} &= u'_{i+2^{m-1},m-1} \nonumber\\
&= u_{j+2^{m-1},m-1} \oplus \left( u_{j,m-1} \cdot \xi_i \right) \nonumber\\
&= \begin{cases}
u_{j+2^{m-1},m-1} = u_{j+2^{m-1},m} & \text{for } \xi_i=0\\
u_{j+2^{m-1},m-1} \oplus u_{j,m-1} = u_{j,m}& \text{for } \xi_i=1\\
\end{cases}\nonumber \\
&= u_{\pi(i+2^{m-1}),m}
\end{align}
which proves Theorem 2. $ \qed $
\end{appendix}

\bibliographystyle{IEEEtran}
\bibliography{references}

\vspace{-1.1cm}

\begin{IEEEbiography}[{\includegraphics[width=1in,height=1.25in,clip,keepaspectratio]{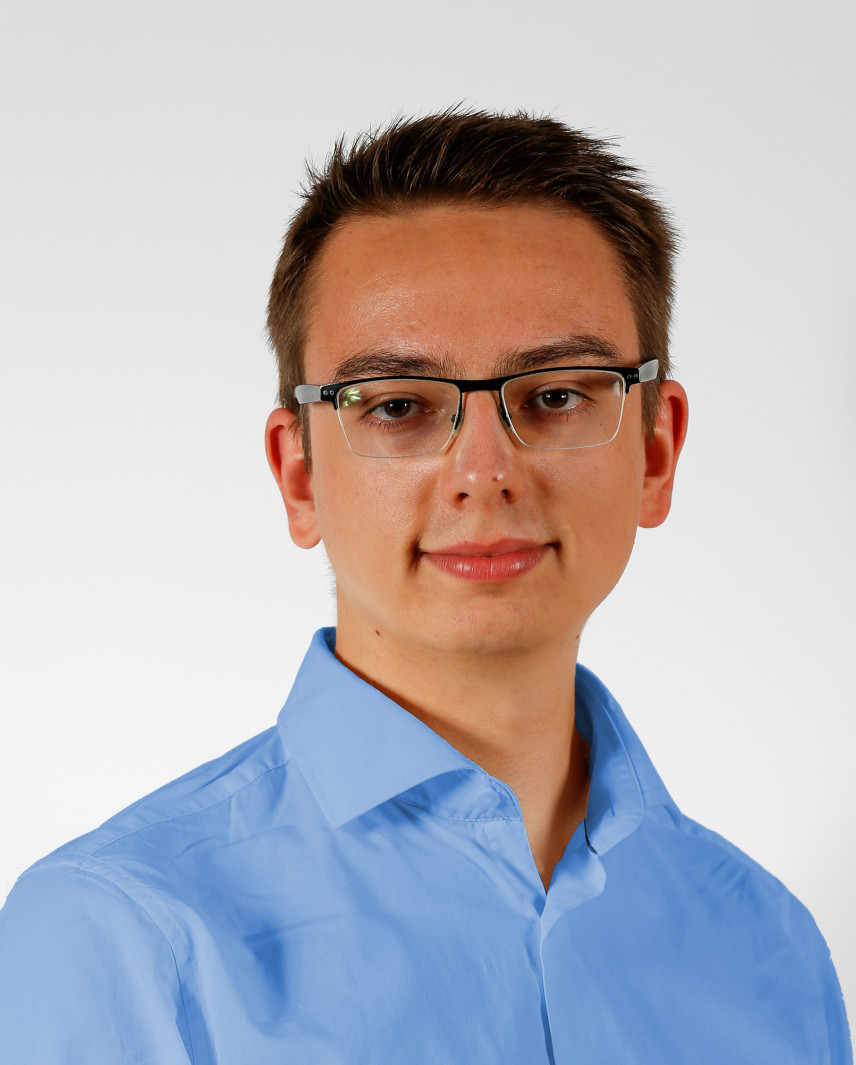}}]
	{Marvin Geiselhart} (S’20) received the B.Sc. and M.Sc. degree (with  distinction) in electrical engineering and information technology from the University of Stuttgart, Germany, in 2017 and 2019, respectively. 
	During his master studies, he worked at Bosch as an intern and working student. 
	He has been a member of the research staff at the Institute of Telecommunications, University of Stuttgart since the beginning of 2020, where he is currently pursuing the Ph.D. degree. 
	His main research topic is channel coding, particularly polar coding and algebraic coding for low-latency applications. 
	He was awarded the Anton- und Klara Röser Preis and the VDE-Preis for his master thesis.
\end{IEEEbiography}

\vspace{-1.1cm}

\begin{IEEEbiography}[{\includegraphics[width=1in,height=1.25in,clip,keepaspectratio]{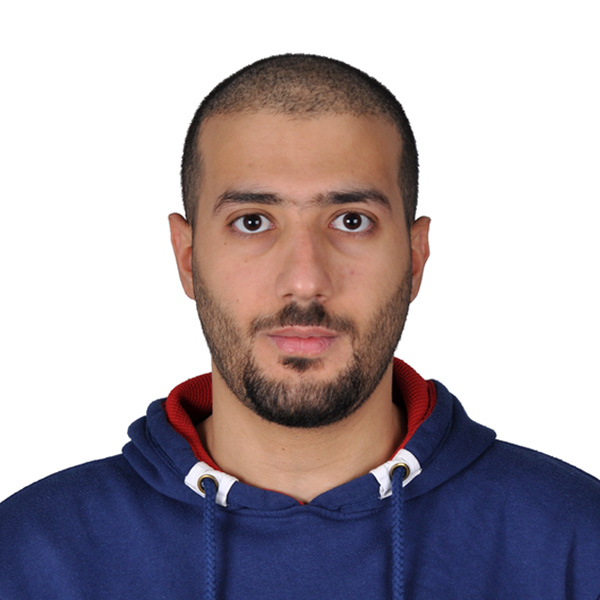}}]
	{Ahmed Elkelesh} received the B.Sc. degree (with highest honours) in Information Engineering and Technology in 2013 from the German University in Cairo and M.Sc. degree (with distinction) in Communications Engineering and Media Technology in 2016 from the University of Stuttgart. 
	During his years of study in Germany, he was a research assistant with Fraunhofer IPA Stuttgart and an intern at Sony Stuttgart Technology Center.
	Since 2016, he has been a member of research staff with the Institute of Telecommunications, University of Stuttgart, where he is working toward the Ph.D. degree. 
	His main research topic is channel coding, with particular emphasis on polar codes and LDPC codes.
	Further research interests include the areas of information theory, modulation, machine learning and SDR.
	He was the recipient of the Anton-und-Klara-R{\"o}ser prize 2017 for his master thesis. 
\end{IEEEbiography}

\vspace{-1.1cm}

\begin{IEEEbiography}[{\includegraphics[width=1in,height=1.25in,clip,keepaspectratio]{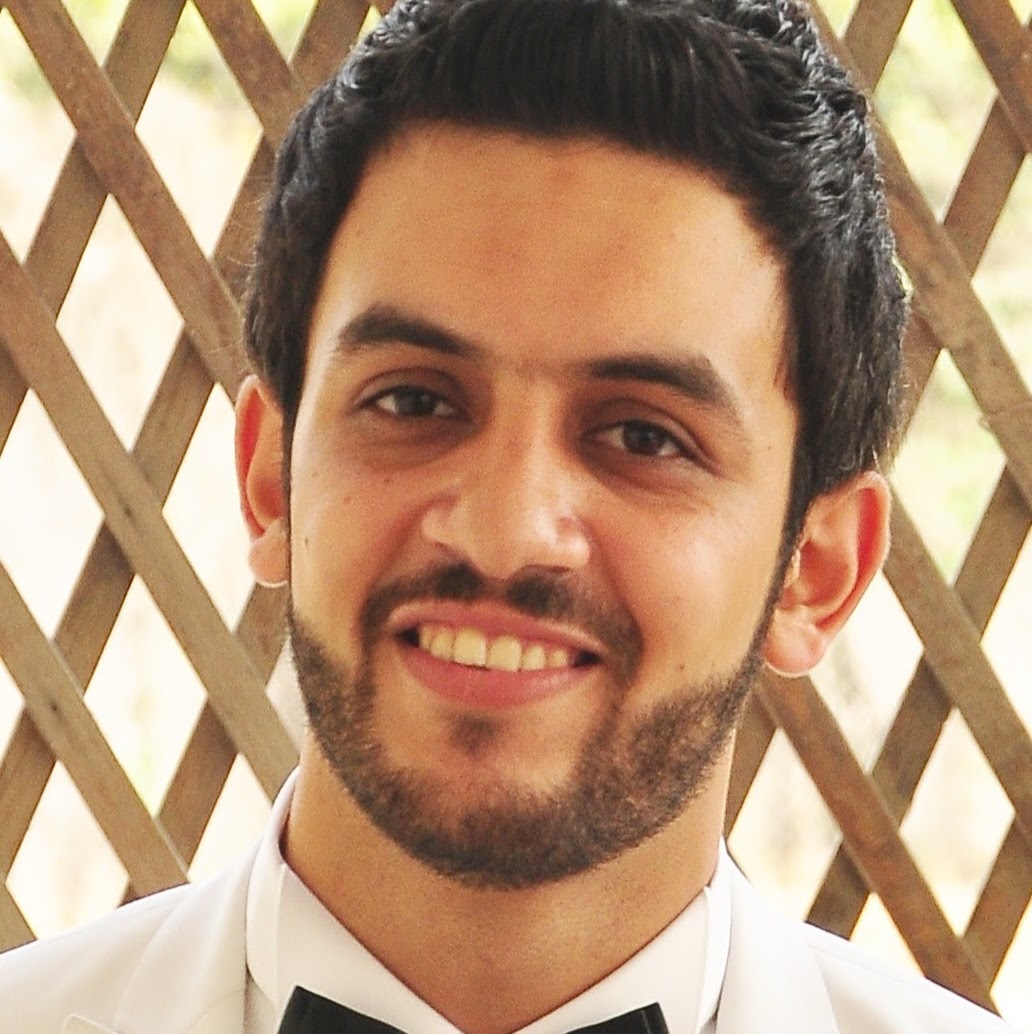}}]
	{Moustafa Ebada} received the B.Sc. (with distinction) from the Communications department in the German University in Cairo, Egypt in 2013 and his M.Sc. degree in electrical engineering and information technology from the University of Stuttgart, Germany in 2016, where he is currently working toward the Ph.D. degree. During his master studies, he was a research assistant with multiple institutes of the University of Stuttgart including fields of radio frequency technology, signal processing and telecommunications. Since 2016, he has been a member of research staff with the Institute of Telecommunications, University of Stuttgart. His main research topics are channel coding, particularly polar code construction and decoding. Besides, designing short LDPC codes for high speed applications. Further research interests include machine learning, particularly designing error correction codes and utilization of the state-of-the-art coding schemes in the field of Multiple Access Channel.  
\end{IEEEbiography}

\vspace{-1.1cm}

\begin{IEEEbiography}[{\includegraphics[width=1in,height=1.25in,clip,keepaspectratio]{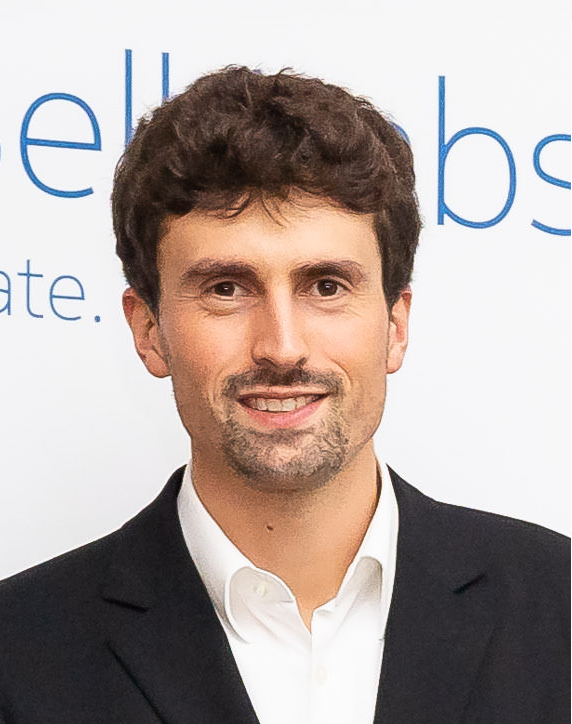}}]
	{Sebastian Cammerer} (S'16) is a research scientist at Nvidia. Before joining Nvidia he received his PhD in electrical engineering and information technology from the University of Stuttgart, Germany, in 2021. His main research topics are machine learning for wireless communications and channel coding. Further research interests are in the areas of modulation, parallelized computing for signal processing and information theory. He is recipient of the IEEE SPS Young Author Best Paper Award 2019, the Best Paper Award of the University of Stuttgart 2018, the Anton- und Klara Röser Preis 2016, the Rohde\&Schwarz Best Bachelor Award 2015, the VDE-Preis 2016 for his master thesis and third prize winner of the Nokia Bell Labs Prize 2019.
\end{IEEEbiography}

\vspace{-1.1cm}

\begin{IEEEbiography}[{\includegraphics[width=1in,height=1.25in,clip,keepaspectratio]{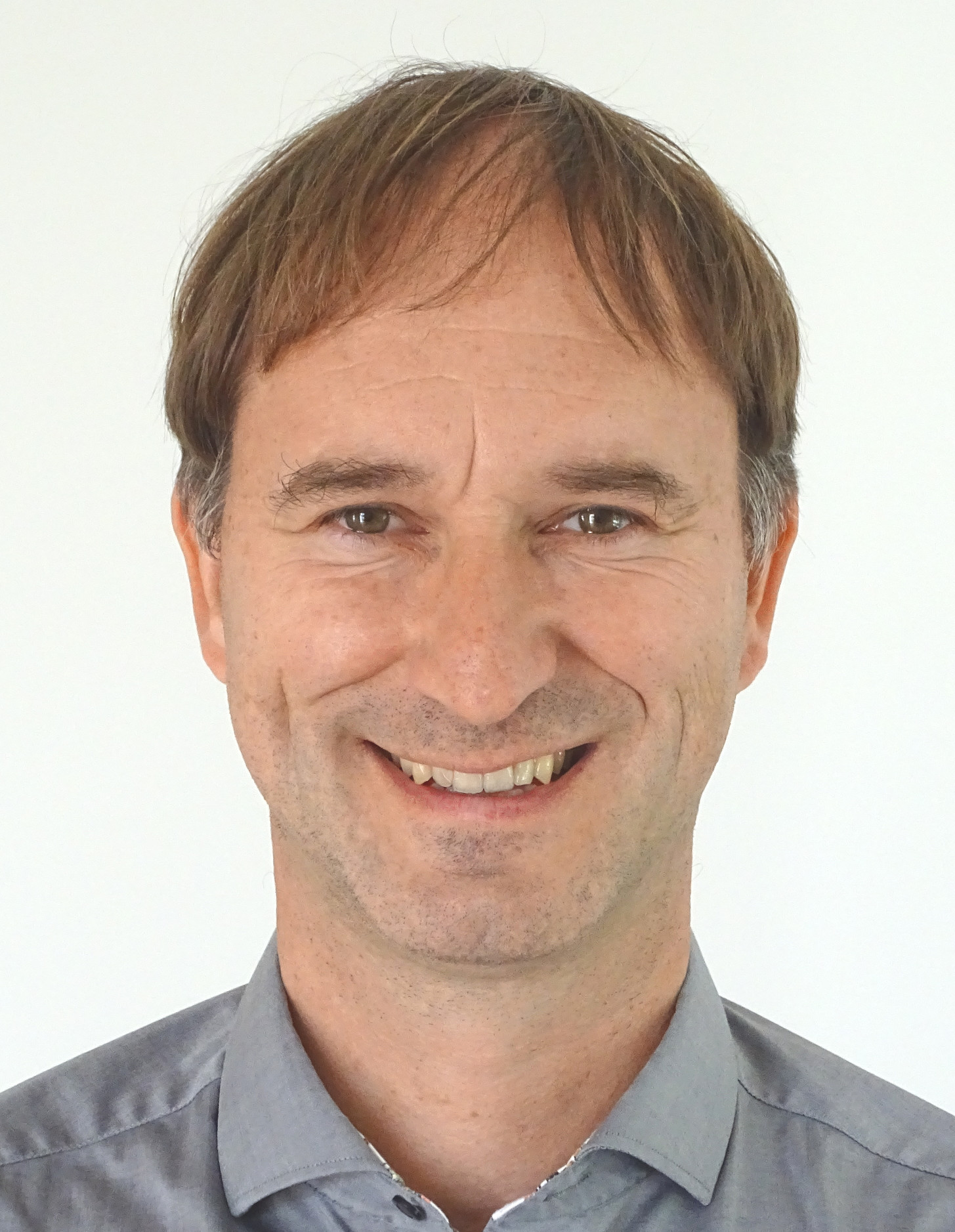}}]
	{Stephan ten Brink}(M'97--SM'11--F'21) has been a faculty member at the 
	University of Stuttgart, Germany, since July 2013, where he is head of 
	the Institute of Telecommunications.
	From 1995 to 1997 and 2000 to 2003, Dr. ten Brink was with Bell 
	Laboratories in Holmdel, New Jersey, conducting research on multiple 
	antenna systems.
	From July 2003 to March 2010, he was with Realtek Semiconductor Corp., 
	Irvine, California, as Director of the wireless ASIC department, 
	developing WLAN and UWB single chip MAC/PHY CMOS solutions.
	In April 2010 he returned to Bell Laboratories as Department Head of the 
	Wireless Physical Layer Research Department in Stuttgart, Germany.
	Dr. ten Brink is a recipient and co-recipient of several awards, 
	including the Vodafone Innovation Award, the IEEE Stephen O. Rice Paper Prize, 
	the IEEE Communications Society Leonard G. Abraham Prize for contributions to channel coding and 
	signal detection for multiple-antenna systems.
	He is best known for his work on iterative decoding (EXIT charts) and 
	MIMO communications (soft sphere detection, massive MIMO).
\end{IEEEbiography}

\end{NoHyper}
\end{document}